\documentclass{aa}
\usepackage{soul}

\usepackage{graphicx}
\usepackage{txfonts}
\usepackage[]{hyperref}
\usepackage{xcolor}
\usepackage{amstext}

\begin{document}

\title{Augmenting photometric redshift estimates using spectroscopic nearest neighbours}

\titlerunning{Augmenting redshift estimates with GNN}

\author{F. Tosone\inst{1}\fnmsep\thanks{\email{\href{mailto:federico.tosone@unimi.it}{federico.tosone@unimi.it}}},
    M.S.~Cagliari\inst{1}\fnmsep\thanks{\email{\href{mailto:marina.cagliari@unimi.it}{marina.cagliari@unimi.it}}},
    L.~Guzzo\inst{1,2,3}, B.R.~Granett\inst{2}, A.~Crespi\inst{1}
    }

\institute{
            Dipartimento di Fisica ``Aldo Pontremoli'', Università degli Studi di Milano, via G. Celoria 16, I-20133 Milano, Italy
            \and
            INAF-Osservatorio Astronomico di Brera, via Brera 28, I-20121 Milano, and via E. Bianchi 46, I-23807, Merate, Italy
            \and
            INFN-Sezione di Milano, via G. Celoria 16, I-20133 Milano, Italy}

\abstract
{As a consequence of galaxy clustering, close galaxies observed on the plane of the sky should be spatially correlated with a probability that is inversely proportional to their angular separation. In principle, this information can be used to improve photometric redshift estimates when spectroscopic redshifts are available for some of the neighbouring objects. Depending on the depth of the survey, however, this angular correlation is reduced by chance projections. In this work, we implement a deep-learning model to distinguish between apparent and real angular neighbours by solving a classification task. We adopted a graph neural network architecture to tie together photometry, spectroscopy, and the spatial information between neighbouring galaxies. We trained and validated the algorithm on the data of the VIPERS galaxy survey, for which photometric redshifts based on spectral energy distribution are also available. The model yields a confidence level for a pair of galaxies to be real angular neighbours, enabling us to disentangle chance superpositions in a probabilistic way. When objects for which no physical companion can be identified are excluded, all photometric redshift quality metrics improve significantly, confirming that their estimates were of lower quality. For our typical test configuration, the algorithm identifies a subset containing $\sim 75 \%$ high-quality photometric redshifts, for which the dispersion is reduced by as much as $50 \%$ (from $0.08$ to $0.04$), while the fraction of outliers reduces from $3\%$ to $0.8\%$.
Moreover, we show that the spectroscopic redshift of the angular neighbour with the highest detection probability provides an excellent estimate of the redshift of the target galaxy, comparable to or even better than the corresponding template-fitting estimate.
}

\maketitle

\section{Introduction}\label{sec:intro}
Knowledge of galaxy distances is of the utmost importance for cosmology to reconstruct the underlying 3D dark matter distribution that encapsulates key information about the evolution and matter content of the Universe. 
On cosmological scales, the most efficient method for estimating distances is through their cosmological redshift, which directly connects to the standard definitions of distance. Sufficiently precise redshift measurements allow us to test the world model through the redshift-distance relation, coupled with standard rulers and standard candles \citep[e.g.][]{Riess98,Perlmutter98}. 

Over the past 25 years, galaxy clustering measurements from large redshift surveys have been able to quantify the universal expansion and growth histories, pinpointing the value of cosmological parameters to high precision \citep[e.g.][]{Tegmark06,Colless03,Blake11,Torre17,Alam17,Pezzotta17,Bautista21}. Even larger redshift surveys are now ongoing \citep[DESI;][]{DESI16} or are scheduled to start soon \citep[Euclid;][]{Laureijs}, with the goal of further refining these measurements to exquisite precision and find clues for the poorly understood ingredients of the remarkably successful standard model of cosmology.

The redshift is measured from the shift in the position of emission and absorption features identified in galaxy spectra, typically through cross-correlation techniques with reference templates, which capture the full available information \citep[e.g.][]{Tonry79}. Despite the considerable advances of multi-object spectrographs over the past 40 years, collecting spectra for large samples of galaxies remains an expensive task. A cheaper, lower-precision alternative is offered by photometric estimates, that is, by measurements based on multi-band imaging, in which integrated low-resolution spectral information is collected at once for large numbers of objects over large areas. The price to be paid is that of larger measurement errors, together with a number of catastrophic failures, which limit the scientific usage of such {\sl photometric redshifts} (photo-zs hereafter) to specific applications \citep[e.g.][]{Newman22}. Still, when a sufficient number of photometric bands is available \citep{jpas,cosmos,pau} or when even information about the ensemble mean spectrum can be obtained \citep{Cagliari2022}, these samples become highly valuable in many respects. 
\begin{figure*}
\centering
\includegraphics[width=1. \textwidth]{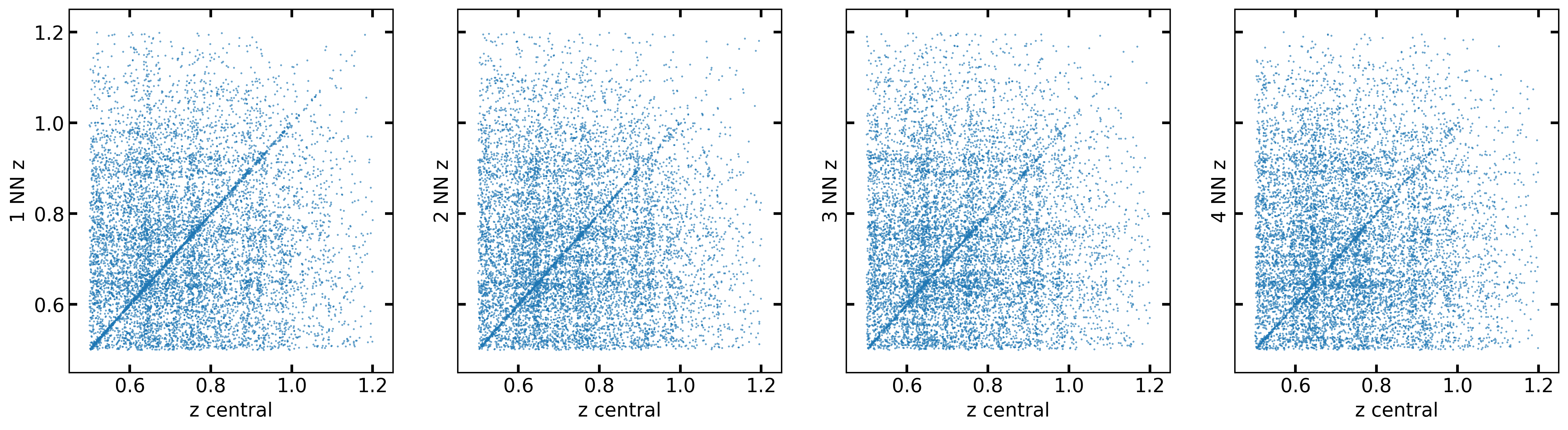}
\caption{Correlation between the galaxy redshift and that of its $nth$ nearest angular neighbour ($n=\{1,2,3,4\}$, left to right), as seen in the VIPERS redshift survey data, which cover the range $0.5< z < 1.2$. Clearly, while a tight correlation exists for a number of objects, many other angular pairs just correspond to chance superpositions. 
\label{fig:NN}}
\end{figure*}
Photo-zs are traditionally estimated by fitting template spectral energy distributions (SED) to the measured photometric fluxes \citep[see e.g.][]{Bolzonella00,Arnouts03,Maraston05,Ilbert06}. Detailed reviews can be found in \citet{Salvato19}, \citet{Brescia21}, and \citet{Newman22}.

Since the pioneering work of \citet[][see also \citeauthor{Lahav94} \citeyear{Lahav94}]{Collister04}, who first used artificial neural networks (ANN) to obtain photo-z estimates, machine-learning (ML) algorithms have seen many further applications in this context. These include {\sl random forests} \citep{Carliles10}, {\sl self-organizing maps} \citep[SOM;][]{Masters15}, and advanced ANNs \citep{Sadeh16}. A notable recent application uses the full images of galaxies through {\sl convolutional neural networks} \citep[CNN;][]{Pasquet19,Henghes22}. All these methods provide photo-z estimates by using information that is strictly local, that is, the flux of each object measured in a number of photometric bands, independently of correlations with the other galaxies in the sample.

In the specific case when a photometric survey includes spectroscopic redshifts for a representative sub-sample spread over the same area, these represent additional information, which can be exploited to obtain improved estimates of the missing redshifts. Since galaxies are spatially clustered, angular neighbours on the sky preserve a degree of redshift correlation, depending on the depth of the catalogue. The deeper the catalogue, the weaker the correlation because the projection is made over a deeper baseline. Still, an angular correlation remains, as can be seen explicitly in Fig. \ref{fig:NN}, in the data of the VIMOS Public Extragalactic Redshift Survey \citep[VIPERS;][]{Guzzo14}.

This correlation was exploited, for example, to improve our knowledge of the overall sample redshift distribution \citep{Newman08}, which is a fundamental quantity for many cosmological investigations such as weak-lensing tomography. With VIPERS, instead, it was used to estimate the galaxy density field to fill the gaps due to missing redshifts  \citep{Cucciati2014}. Even more finely, \citet{Aragon-Calvo15} used the fact that galaxies are typically confined within cosmic web structures to obtain a dramatic improvement in the estimate of photo-zs for $\sim 200$ million Sloan Digital Sky Survey galaxies, starting from only about one million spectroscopically measured redshifts. 

Our goal with the work presented here has been to optimally retrieve this non-local information from the neighbouring objects of a given galaxy building upon a specific class of ML architectures, graph neural networks (GNN). The key property of this class is the ability to combine information from unstructured data based on our priors of the task at hand \citep{Bronstein17}. The end goal is to obtain an improved estimate of the galaxy redshift.

As shown by Fig. \ref{fig:NN}, the existing correlation between angular neighbours is strongly diluted by the sea of chance superpositions along the line of sight. Thus, the problem can be more appropriately recast into quantifying the probability that a given angular neighbour (with known redshift) is a physical companion for a given galaxy and thus is closely correlated in redshift as well. 
Our GNN model, dubbed {\sl NezNet}, combines the intrinsic features of a {\sl target} galaxy and a {\sl neighbour}, that is, their multi-band fluxes, the spectroscopic redshift of the neighbour, and  their relative angular distance, to output the probability for the two galaxies to be spatially correlated. We trained and tested NezNet using the spectroscopic sample of VIPERS. We show that discarding targets for which no real physical neighbour is identified with significant probability improves the quality of the associated photo-z catalogue obtained through classic SED fitting, increasing precision and accuracy and reducing the fraction of catastrophic outliers. Moreover, when real neighbours are identified, the redshift of the highest-probability neighbour represents an estimate of the target redshift that is typically more precise than that obtained through the classical SED fitting.

The idea of using GNNs to draw additional redshift information from neighbouring galaxies is not new. \citet{Beck2019} presented preliminary results of an approach based on using only the photometry of a neighbourhood of galaxies, obtaining a 10\% improvement on the median absolute deviation of the photo-zs estimated via a single object-based ML algorithm. The main shortcoming of methods that are based on apparent neighbours lies in the large fraction of chance superpositions, as evident in Fig. \ref{fig:NN}. Here, we reformulated the problem as a detection task that identifies the physical neighbours of the surrounding spectroscopic objects, also including the neighbour's spectroscopic information. In this way, we obtain a significant improvement.

The paper is organised as follow. In Sect. \ref{sec:model} we give a brief description of how GNNs work and specify the architecture of our model. In Sect. \ref{sec:data} we describe the properties of VIPERS data and the way we prepared the training set, in particular, how we defined real or apparent neighbouring objects. Section \ref{sec:method} describes how the model is applied to the data and the metrics we used to quantify the performance of the results. Finally, in Sect. \ref{sec:results} we present and discuss our results, and we conclude in Sect. \ref{sec:conclusions}. 

\section{Model}\label{sec:model}
\begin{figure*}%[ht!]
\centering
\includegraphics[width=.9\textwidth]{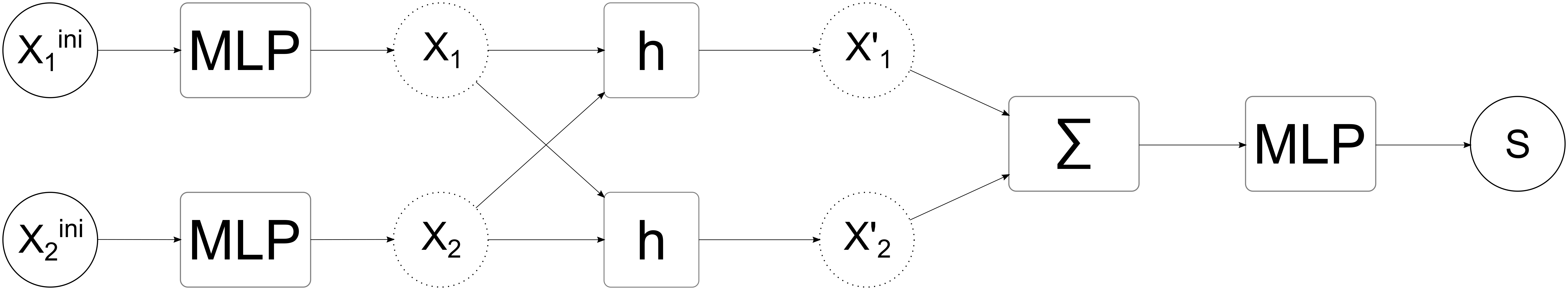}
\caption{Schematic architecture of NezNet. The input features are first processed by a dense network. Message passing between the two layers through Eq. (\ref{eq:EdgeConv}) is then applied to take the relative differences and global values of the features into account. Before the final dense layer, the features are summed and then reprocessed with an MLP to output the score probability of two galaxies being actual neighbours.}\label{fig:model}
\end{figure*}
A neural network model can be summarised as a set of non-linear functions applied to a set of inputs that undergo a linear mapping. Each mapping has many parameters that are optimised through a training process that allows the network model to approximate a wide variety of almost arbitrary functions \citep{LeCun15}.  In its simplest form, a neural network model corresponds to a multi-layer perceptron (MLP), also known as dense neural network \citep{Murtagh91}. For images, neural architectures such as CNN are more suited because they take our a priori knowledge about the data structure into account \citep{OShea15}.

This reasoning can be pushed further by introducing neural networks for graph representations \citep{Zhou18}. In this work, we make use of one key aspect of GNN, that is, message passing \citep{Gilmer17}. To fix ideas, the problem we wish to address is the following: we need to find the spectroscopic galaxies with the highest probability of being close to a galaxy for which only photometric information is available. This can be recast as a classification task for each pair of galaxies, in which our aim is to distinguish between apparent and real neighbours when projected on the plane of the sky.

Intuitively, a model that distinguishes between apparent and real neighbours should be based on the relative difference between galaxy features. A neural network like this can be designed by including a layer of the form
\begin{equation}\label{eq:EdgeConv}
    \pmb{x}'_i = \underset{j \in \mathcal{N}(i)}{\sum} h(\pmb{x}_i, \pmb{x}_i - \pmb{x}_j) \, ,
\end{equation}
where $\pmb{x}_i$ refers to the array of input features of the node $i$, $\mathcal{N}(i)$ is the neighbourhood of the same node, $\sum$ is the aggregation function that sums the outcomes from each pair of nodes. The function $h$ is an MLP that explicitly combines the value of the input feature at the node and the relative difference of that feature with respect to the neighbour. It is worth noting that this GNN is both permutation equivariant and permutation invariant, so that it is not affected by a change in the order of the nodes, that is, the input galaxies.

The complete architecture of our model is illustrated in Fig. \ref{fig:model}. Each node is a galaxy, whose inputs (e.g. the photometric measurements) were pre-processed through an MLP before undergoing the message passing of Eq. (\ref{eq:EdgeConv}). We restricted ourselves to the case of galaxy pairs, so that the neighbourhood $\mathcal{N}(j)$ includes only one galaxy, and the aggregation function simply sums the features $\pmb{x}'_{1}+\pmb{x}'_{2}$. This model can be seen as a trivial version of EdgeConv \citep{Wang18}, where the adjacency matrix is a $2 \times 2$ matrix, with $0$ entries for diagonal elements and $1$ for the off-diagonal elements. Finally, the summed features undergo a last dense layer with a scalar output. All the activation functions are rectified linear units, with the exception of the final layer, where we used a sigmoid, to represent a probability for our classification task.

We call this classification model Nearest-z Network (NezNet). NezNet provides the probability for a pair of galaxies to be real neighbours. The loss function adopted to train NezNet is a standard binary cross entropy,
\begin{equation}\label{eq:loss}
    \mathcal{L} =  \frac{1}{n} \sum^{n}_{i} \left[ y_i \log{p_i} + (1-y_i) \log{\left(1-p_i\right)} \right],
\end{equation}
where $p_i$ is the output probability of NezNet for each galaxy pair, while $y_i={0,1}$ is the corresponding training label, and the sum is averaged over the mini-batch. To design our model, we made use of the Spektral library\footnote{\url{https://graphneural.network}} \citep{Grattarola20}, where the EdgeConv layer is conveniently already implemented. 

\section{Data}\label{sec:data}
We trained and tested our approach on the final data release of VIPERS \citep{Guzzo14,Scodeggio17}, for which the redshift correlation between angular neighbours is shown in Fig. \ref{fig:NN}. 
The survey used the VIMOS multi-object spectrograph at the ESO Very Large Telescope to target galaxies brighter than $i_{\rm AB}=22.5$ in the Canada-France-Hawaii Telescope Legacy Survey Wide (CFHTLS-Wide) catalogue, with an additional $(r-i)$ vs $(u-g)$ colour pre-selection to remove objects at $z<0.5$. The resulting sample covers the redshift range $0.5 \lesssim z \lesssim 1.2$, with an effective sky coverage of $16.3 \, \mathrm{deg}^2$, split over the W1 and W4 fields of CFHTLS-Wide. We used only galaxies with secure redshift measurements, as identified by their quality flag, corresponding to a $96.1 \%$ confidence level (see \citeauthor{Scodeggio17} \citeyear{Scodeggio17}).

For each galaxy in the catalogue, the following information was considered: the spectroscopic redshift measurement $z_{\text{spec}}$, the six magnitudes $u$, $g$, $r$, $i$, $z$ (not to be confused with redshift) and $K_{\mathrm{s}}$, the right ascension $\alpha$ ($\mathrm{RA}$), in radians, and the declination $\delta$ ($\mathrm{Dec}$), in radians.

The angular separation on the sky between two objects with RA $\alpha_1$ and $\alpha_2$ and Dec $\delta_1$ and $\delta_2$ is given by the haversine formula,
\begin{equation}\label{eq:angular_distance}
    \Delta \Theta = \arccos{\left( \sin{\delta_1} \sin{\delta_2} + \cos{\delta_1} \cos{\delta_2} \cos{\left(\alpha_1-\alpha_2\right)} \right).}
\end{equation}

We selected the parent photometric sample by applying the same VIPERS colour and magnitude cuts defined above, so as to be fully coherent with the spectroscopic data. 

\section{Application}\label{sec:method}
We set up a training set from the VIPERS W1 galaxy catalogue. We randomly selected about $3\cdot 10^4$ target galaxies, whose spectroscopic redshift during training was ignored. 
For each of them, we identified the first $n_{\text{NN}}$ angular nearest neighbours as defined by Eq. (\ref{eq:angular_distance}), which we called spectroscopic galaxies because their spectroscopic redshift information was used in our model. Each of these spectroscopic neighbours was associated with the same target galaxy, but the pairs can be considered as independent from one another in our model. Each angular pair was assigned label $1$ when it was a real physical pair, otherwise, it was assigned a $0$. The training set was thus made of galaxy pairs.

A target galaxy of a pair can also be the nearest neighbour of another target galaxy in another pair. We made this choice in order to maximise the number of training examples available in W1. Our final tests on the W4 catalogue show that this does not lead to any over-fitting of VIPERS data, as the model generalises well. We note that this setting assumes a ratio of spectroscopic to photometric objects of $1:1$. In the Conclusions section (Sect. \ref{sec:conclusions}), we also confirm these results in the more realistic case in which the number of spectroscopic redshifts used for training are a fraction of the number of photometric objects.

The definition of a real neighbour is arbitrary; it is reasonable to consider that two angular neighbours form a physical pair when their spectroscopic separation is smaller than a given threshold,
\begin{equation}\label{eq:threshold}
    \Delta z \, (1+z_{\text{spec}}).
\end{equation}
This means that in setting up the training data, there are two hyper-parameters, the number of nearest neighbours $n_{\text{NN}}$ to be considered, and the spectroscopic separation $\Delta z$. As we show below, these two hyper-parameters can affect the results significantly, and it is thus relevant to set them up wisely, depending on the specific survey.

For each galaxy in the pairs, the input features of the nodes in NezNet are the photometry, the spectroscopy, and the angular position, as listed in Sect. \ref{sec:data}. For the target galaxy, we always set $z_{\text{spec}}=0$, so that the model considered it as a missing feature, while providing its value for the neighbouring galaxy. Magnitudes were normalised to the range $[0,1]$, as computed over the whole VIPERS dataset. The angular inputs were provided in terms of relative distance with respect to the target galaxy, so that $\Delta \Theta=0$ for the latter, while for the neighbour, it  corresponded to Eq. (\ref{eq:angular_distance}). By adopting this choice, we guaranteed that the model has translational invariance. 

Another tested option (see Sect. \ref{sec:conclusions}) is to use the relative distance in the two sky coordinates $\mathrm{RA}$ and $\mathrm{Dec}$ as input variables instead of the angular separation of the two galaxies. This choice arises because the surface distribution of the sample is not rotationally invariant on the sky because of the technical set-up of the slits in the VIMOS focal plane, with the spectral dispersion oriented along the declination direction. As spectra must not overlap on the detector, targets need to be separated in Dec much more than in RA. As a result, the minimum separation is $\sim 1.9 \, \mathrm{arcmin}$ in Dec and $5 \, \mathrm{arcsec}$ in RA. More details can be found in \citet{Bottini05} and \citet[][see their Sect. 4.1]{Pezzotta17}.
Our experiments show that providing the model with the angular separation $\Delta \Theta$ introduces a bias in the redshift metrics, which is not observed when the relative separations along RA and Dec are given. In general, however, we find that the separation information does not significantly improve the classifier, and for this reason, we did not use it in our final model. Spatial information instead comes only from the number of nearest neighbours considered. 

The other hyperparameters of the model, that is, the batch size, number of neurons, and learning rate, have a far weaker impact than $\Delta z$ and $n_{\text{NN}}$, and were set to fiducial values: a batch size of $32$, a learning rate of $0.001$, and a total number of parameters of the order of a few thousands. We find little difference in the output metrics of the redshift estimates when the complexity of the model is increased, or when the batch size and the learning rate are changed around these fiducial values.

NezNet gives as output the probability for two galaxies to be real neighbours. As each target galaxy corresponds to $n_{\text{NN}}$ independent pairs, we can select the neighbour with the highest probability among them. If this probability is below the classification threshold set to define a positive case, we conclude that there is no physical neighbour for that target galaxy in the catalogue. This implies that the probability for the latter is too high to be an outlier in terms of its properties when compared to its neighbours. Removing these objects from the final catalogue significantly improves the metrics when comparing photo-z and spectroscopic measurements. In particular, the reduction in the number of catastrophic redshifts confirms our assumption.  Finding a true neighbour instead reinforces the confidence in the photo-z. At the same time, the spectroscopic redshift of the neighbour in this case is typically an even better estimate of the target redshift than the SED-estimated photo-z . These tests are discussed in the following section.
 
The quantitative comparison between NezNet results,   spectroscopic measurements $z_{\text{spec}}^{(i)}$ , and SED-fitting estimated photo-zs was performed using the metrics defined in \cite{Salvato19}. These are the precision (i.e. the dispersion of the estimated values),
\begin{equation}\label{eq:precision}
    \sigma = \sqrt{\frac{1}{N}\sum_{i}^{N} \left( \frac{z^{(i)}_{\text{spec}} - z^{(i)}}{1+z^{(i)}_{\text{spec}}} \right)^2} \, ,
\end{equation}
the bias
\begin{equation}\label{eq:bias}
    b = \frac{1}{N}\sum_{i}^{N} ( z^{(i)}_{\text{spec}} - z^{(i)} ) \, ,
\end{equation}
and the absolute bias
\begin{equation}\label{eq:absbias}
    |b| = \frac{1}{N}\sum_{i}^{N} |z^{(i)}_{\text{spec}} - z^{(i)} | \, ,
\end{equation}
quantifying systematic deviations. Finally, the outliers are defined as objects for which 
\begin{equation}\label{eq:out}
    |z^{(i)}_{\text{spec}} - z^{(i)}| \geq 0.15 (1+z^{(i)}_{\text{spec}}) \, .
\end{equation}
All the results presented in the following section were obtained by applying the trained NezNet to a test catalogue built in a similar fashion to W1, randomly selecting about $2\cdot10^4$ galaxies from the twin W4 field of VIPERS.

Finally, in the following discussion about our classifier, we use the notion of the true positive rate (TPR), which is the fraction of correctly predicted positive examples with respect to all the real positive examples. It is defined as
\begin{equation}\label{eq:TPR}
    {\rm TPR} = \frac{TP}{TP + FN} \, ,
\end{equation}
where $TP$ stands for true positives and $FN$ stands for false negatives. Similarly, we can define the false positive rate (FPR), which is the fraction of negative examples classified as positives with respect to all the real negative examples, which reads
\begin{equation}\label{eq:FPR}
    {\rm FPR} = \frac{FP}{FP + TN} \, ,
\end{equation}
where $FP$ stands for false positives and $TN$ stands for true negatives.

\section{Results}\label{sec:results}
\begin{figure*}
\centering
\includegraphics[width=1\textwidth]{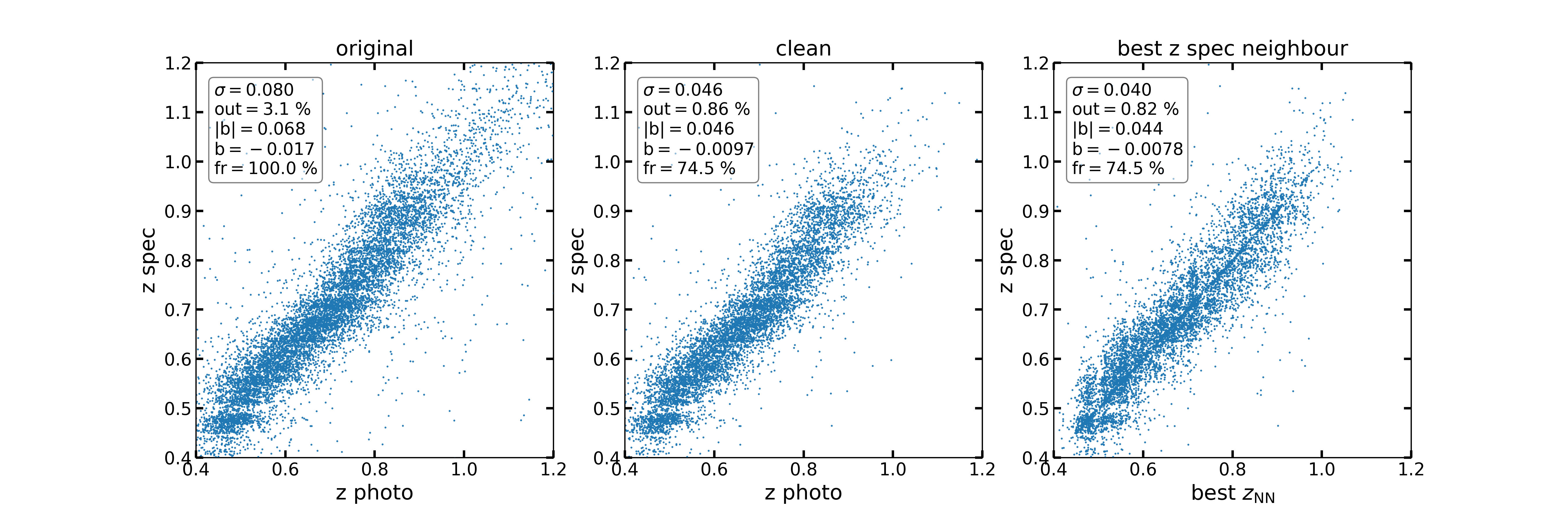}
\caption{Central galaxy spectroscopic redshift versus its photometric redshift measured with and without NezNet. The left panel shows the distribution of photometric vs spectroscopic estimates in the original data. In the middle panel, we show the same distribution after removing the galaxies with low score probability from the catalogue (fr stands for the fraction of retained data). Finally, the right panel shows redshift estimates by assigning the spectroscopic redshift of the neighbour with the highest detection probability to the target galaxy. The model was trained with $n_{\text{NN}}=30$ and $\Delta z = 0.08$.
\label{fig:results_008}}
\end{figure*}
\begin{figure*}%[ht!]
\centering
\includegraphics[width=1\textwidth]{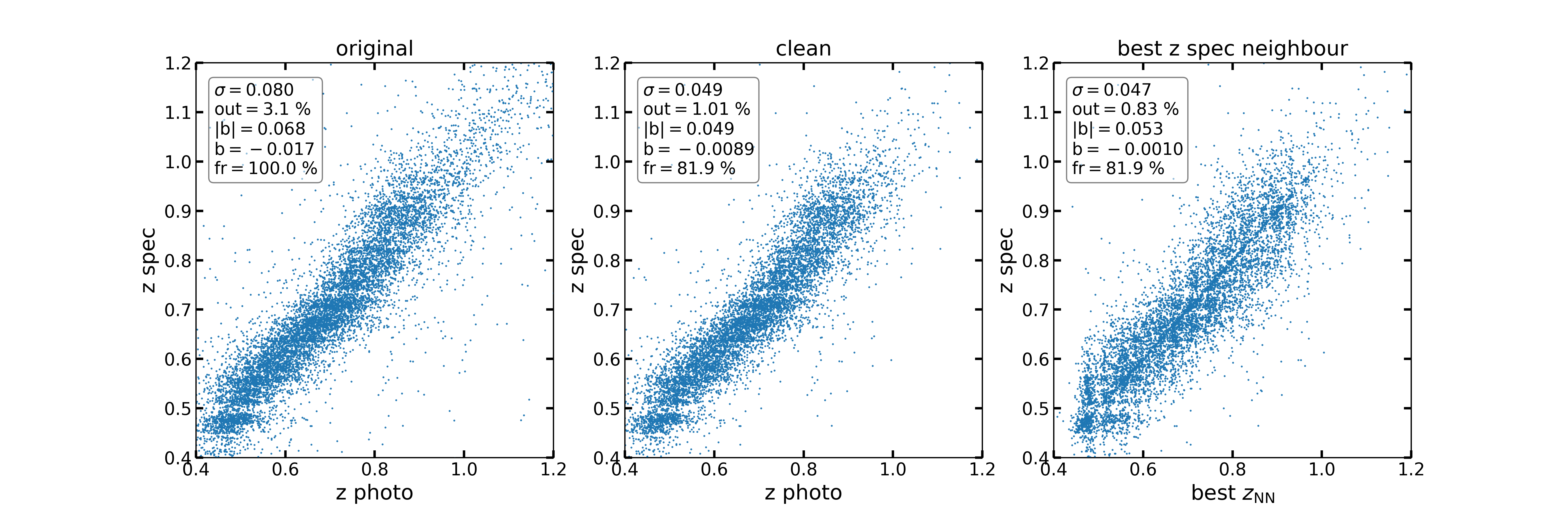}
\caption{Same as Fig. \ref{fig:results_008}, but the model was trained with the higher $\Delta z = 0.15$, while $n_{\text{NN}}=30$ is the same as before. \label{fig:results_015}}
\end{figure*}
\begin{figure*}
\centering
\includegraphics[width=1\textwidth]{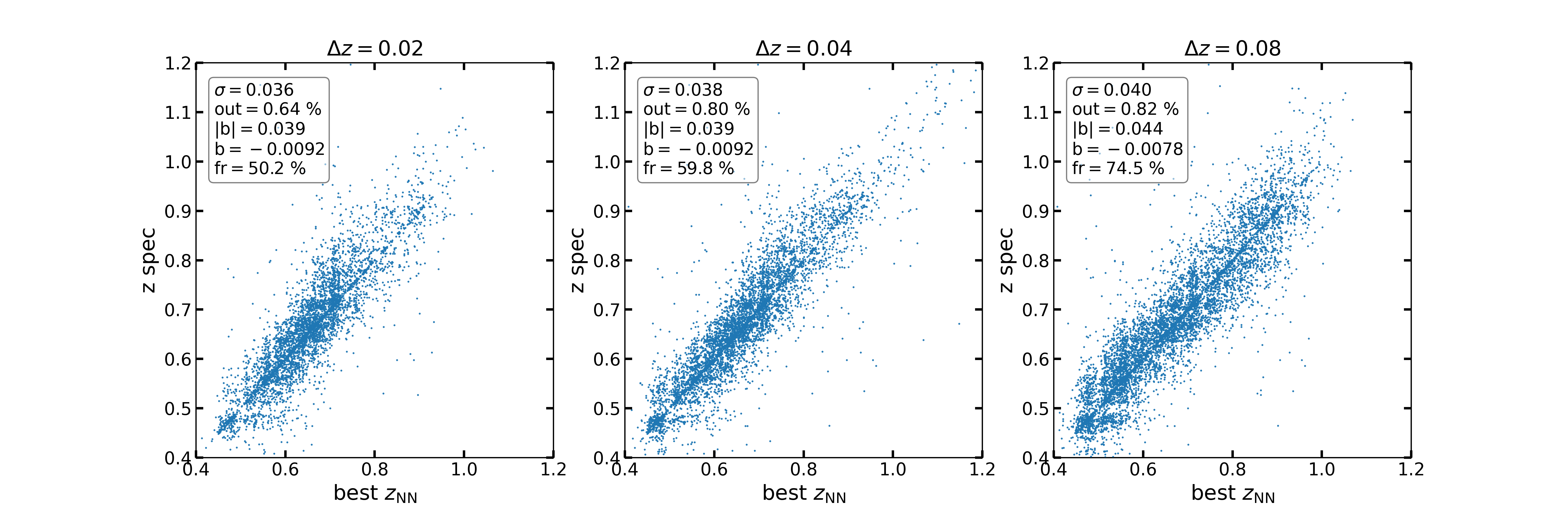}
\caption{Redshift estimates derived from the best nearest neighbour for various $\Delta z$ at fixed $n_{\text{NN}}=30$. Increasing the spectroscopic separation to define physical neighbours while diminishing the quality of the metrics increases the fraction of data that are not dismissed from the catalogue.
\label{fig:results_dz}}
\end{figure*}

As explained in the previous section, NezNet can be used to simply clean a photo-z sample by discarding low-probability neighbours or to provide an alternative redshift estimate derived from the highest-probability neighbour. This is demonstrated on the test catalogue in Fig. \ref{fig:results_008} for a model trained using the hyperparameters $\Delta z=0.08$ and $n_{\text{NN}}=30$. In addition to the VIPERS spectroscopic redshifts, this comparison also includes the original photo-zs estimated by \cite{Moutard16} using standard SED fitting. For these and all following results, angular information (i.e. the separation of the two objects on the sky) was not used as an input variable. The reason for this was already mentioned in the previous section, and is discussed again in more detail below.

Figure \ref{fig:results_008} shows that by simply dismissing the outliers as identified by NezNet, all the metrics improve significantly (central panel). Moreover, when the best neighbour redshifts are adopted for the target galaxies (right panel), we obtain metrics that are comparable to or even better than those of the cleaned photo-z sample. It is worth noting that in this case, the plot shows a characteristic checkerboard pattern because the spectroscopic redshift striping is reflected, as spectroscopic redshifts are now assigned to target photometric objects.

Figure \ref{fig:results_008} also shows the limits of the method. Comparing the left panel with the other two, we can note that NezNet tends to cut off the high-redshift tail of the distribution. This is easily understood considering the magnitude-limited ($i_{\text{AB}}<22.5$) character of the sample used here, which becomes very sparse at $z \gtrsim 1$, where only rare luminous galaxies are present. This means that the model becomes intrinsically less efficient because fewer real physical neighbours are available both for the training and for inference, as is also evident from the density of points at high redshift in Fig. \ref{fig:NN}. Devising a different loss function to up-weight the few physical pairs in this regime might improve the classification task, but an intrinsic limit to the method clearly exists when the density of the sample decreases.

Figure \ref{fig:results_015} shows the same set of plots, but using a higher value for the spectroscopic separation in the training, that is, $\Delta z=0.15$. As expected, allowing for a larger separation in the definition of real angular neighbours discards fewer data. Conversely, there is in general a lower precision and a small increase in the fraction of outliers.

In principle, using a stricter $\Delta z$ could remove even more outliers, retaining only pairs that are closer in redshift and leading to a smaller, but more precise sub-sample. 
We explore this dependence in Fig. \ref{fig:results_dz}. Overall, this method is always able to clean poor estimates from the sample, but at the price of discarding many data points. The minor improvement in precision probably does not justify the use of $\Delta z<0.08$ in the case of VIPERS, because more than half of the sample is excluded.

\begin{figure}
\centering
\includegraphics[width=.55\textwidth]{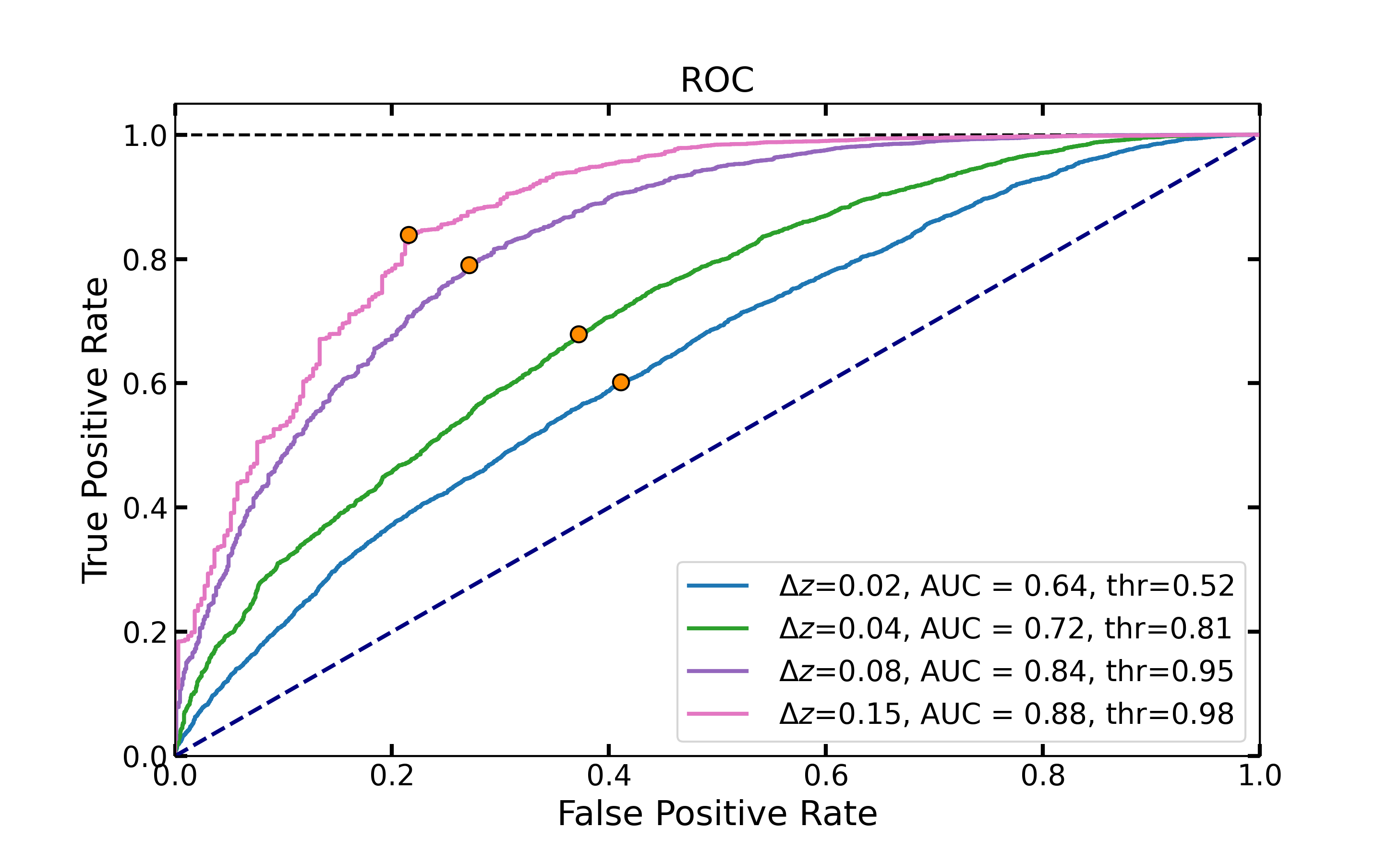}
\caption{ROC curve for a varying redshift threshold $\Delta z$ at fixed $n_{\text{NN}}=30$. The performance of our classifier (AUC) improves as we use a less strict definition of what we define as a true neighbour. The probability that an angular neighbour is a physical neighbour increases at larger $\Delta z$, which is also reflected by the high detection threshold (thr). \label{fig:roc_dz}}
\end{figure}
\begin{figure}
\centering
\includegraphics[width=.55\textwidth]{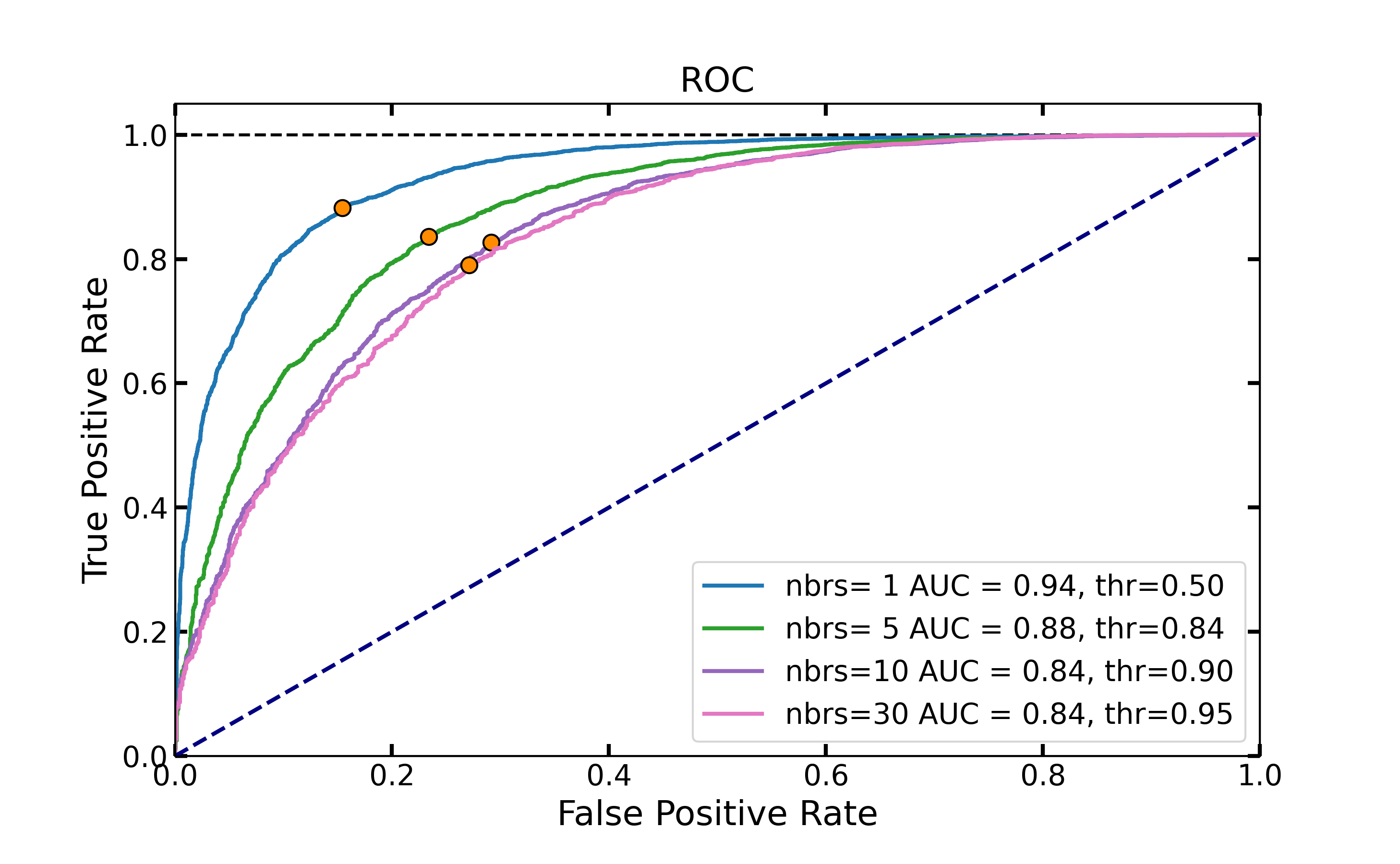}
\caption{ROC curve for a varying number of nearest neighbours $n_{\text{NN}}=30$ at fixed $\Delta z=0.08$. Increasing the number of neighbours that are given in input to the training seems to make the training more difficult. However, this test of the classifier does not reflect the quality of the final redshift estimate, as Fig. \ref{fig:results_nn} shows.
\label{fig:roc_nn}}
\end{figure}
It is apparent that the hyper-parameter $\Delta z$ is very relevant for the quality of the classifier. This is made clear by the receiver operator characteristic (ROC) curve in Fig. \ref{fig:roc_dz}, which shows the TPR (Eq. \ref{eq:TPR}) against the FPR (Eq. \ref{eq:FPR}), and has been computed from the target galaxies in the test catalogue by considering their neighbour with the highest probability. In general, the area under the curve (AUC) is higher for the better classifier. Increasing $\Delta z$ increases the AUC, which would tend to unity for very high values of this parameter, as all galaxies would then be considered real neighbours. However, our ultimate goal is not to increase the performance of the classifier per se, but to improve the metrics of our redshift estimates. These show that $\Delta z \gtrsim 0.08$ represents the best choice for VIPERS.

\begin{figure*}
\centering
\includegraphics[width=1\textwidth]{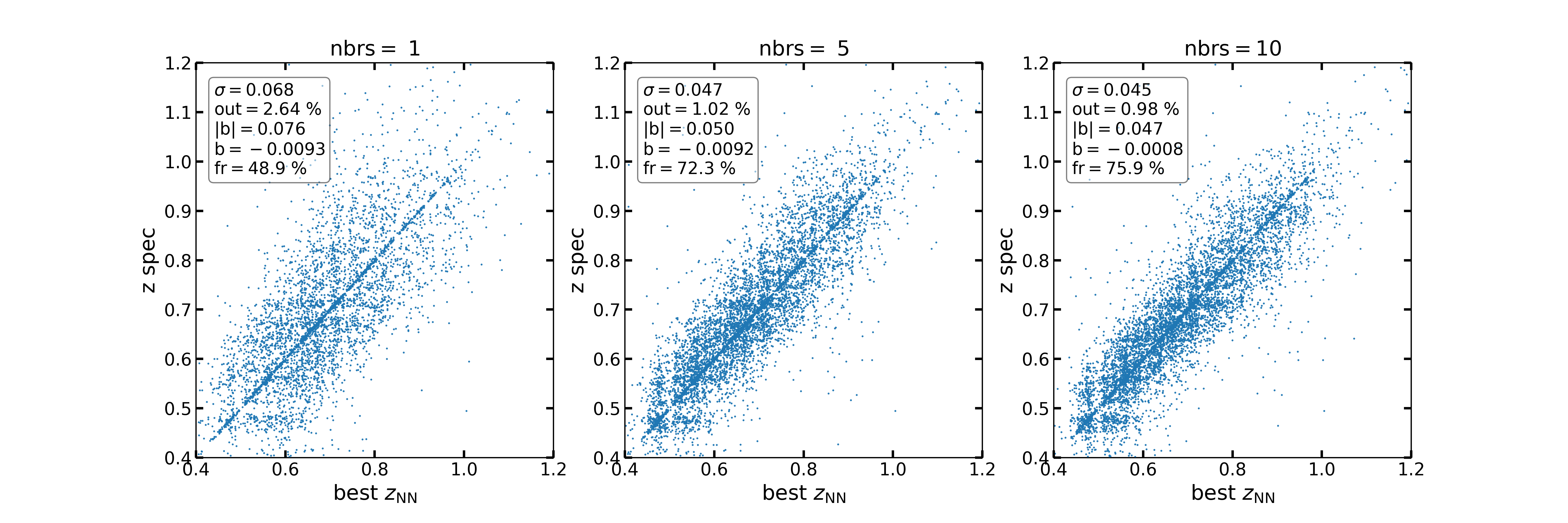}
\caption{Redshift estimates based on the best nearest neighbour for various $n_{\text{NN}}$ at fixed $\Delta z= 0.08$.
Increasing the number of nearest neighbours for each target improves the performance of NezNet in estimating redshifts, as it increases the probability that physical pairs are considered.
\label{fig:results_nn}}
\end{figure*}
The other hyper-parameter of NezNet, that is, $n_{\text{NN}}$, the number of nearest neighbours considered in the training, has a weaker impact on the classifier. We show this in Fig. \ref{fig:roc_nn}, where each  ROC curve corresponds to a model trained with a different $n_{\text{NN}}$, but all with the same $\Delta z$. A drastic change in $n_{\text{NN}}$ does not correspond to comparable changes in the AUC. However, $n_{\text{NN}}$ has a large impact on the redshift estimates, as Fig. \ref{fig:results_nn} shows. A larger number of angular neighbours increases the probability of finding a physical pair, as is shown by the metrics in Fig. \ref{fig:roc_nn}.  
We also experimented with a higher value of $n_{\text{NN}}$ up to 50, but found no further gain with respect to using $n_{\text{NN}}=30$. The redshift metrics start to saturate to the optimal
values already above $n_{\text{NN}}=10$ .

\begin{figure}
\centering
\includegraphics[width=.5\textwidth]{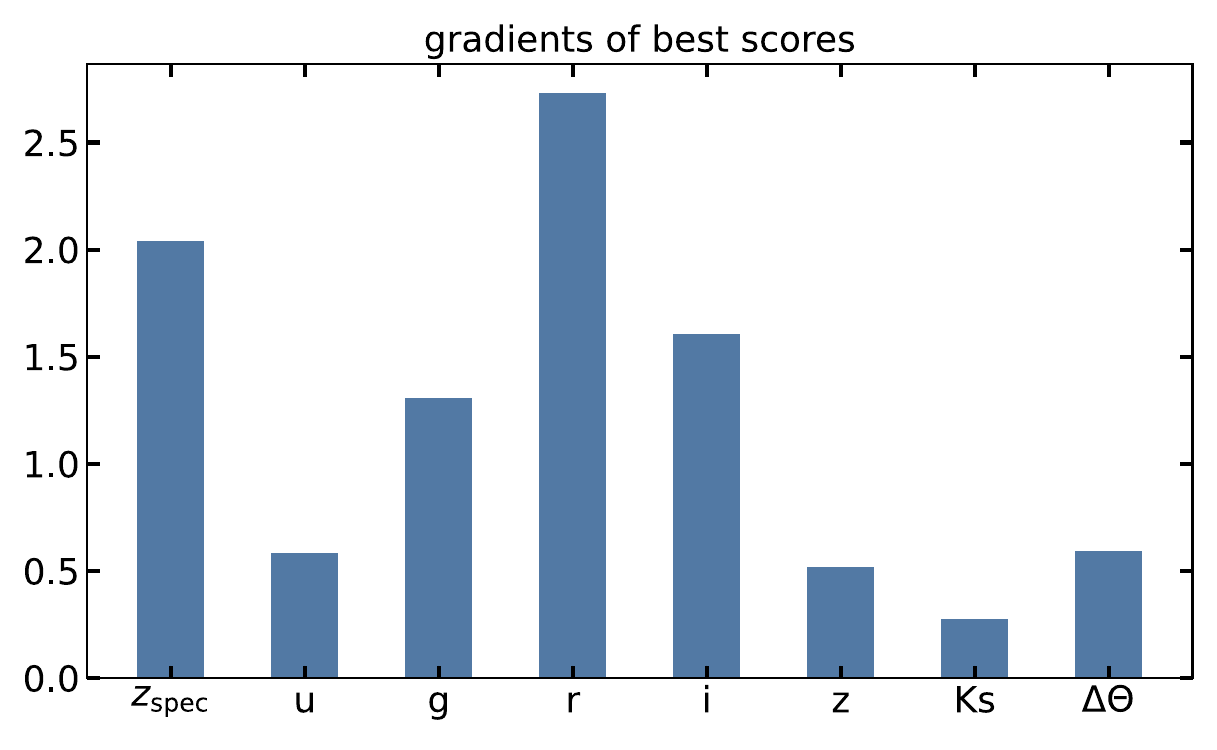}
\caption{Average absolute values of the gradients of NezNet with respect to the input features of the neighbours. For each target, we only considered the neighbour with the highest probability. \label{fig:gradients}}
\end{figure}
\begin{figure}
\centering
\includegraphics[width=.5\textwidth]{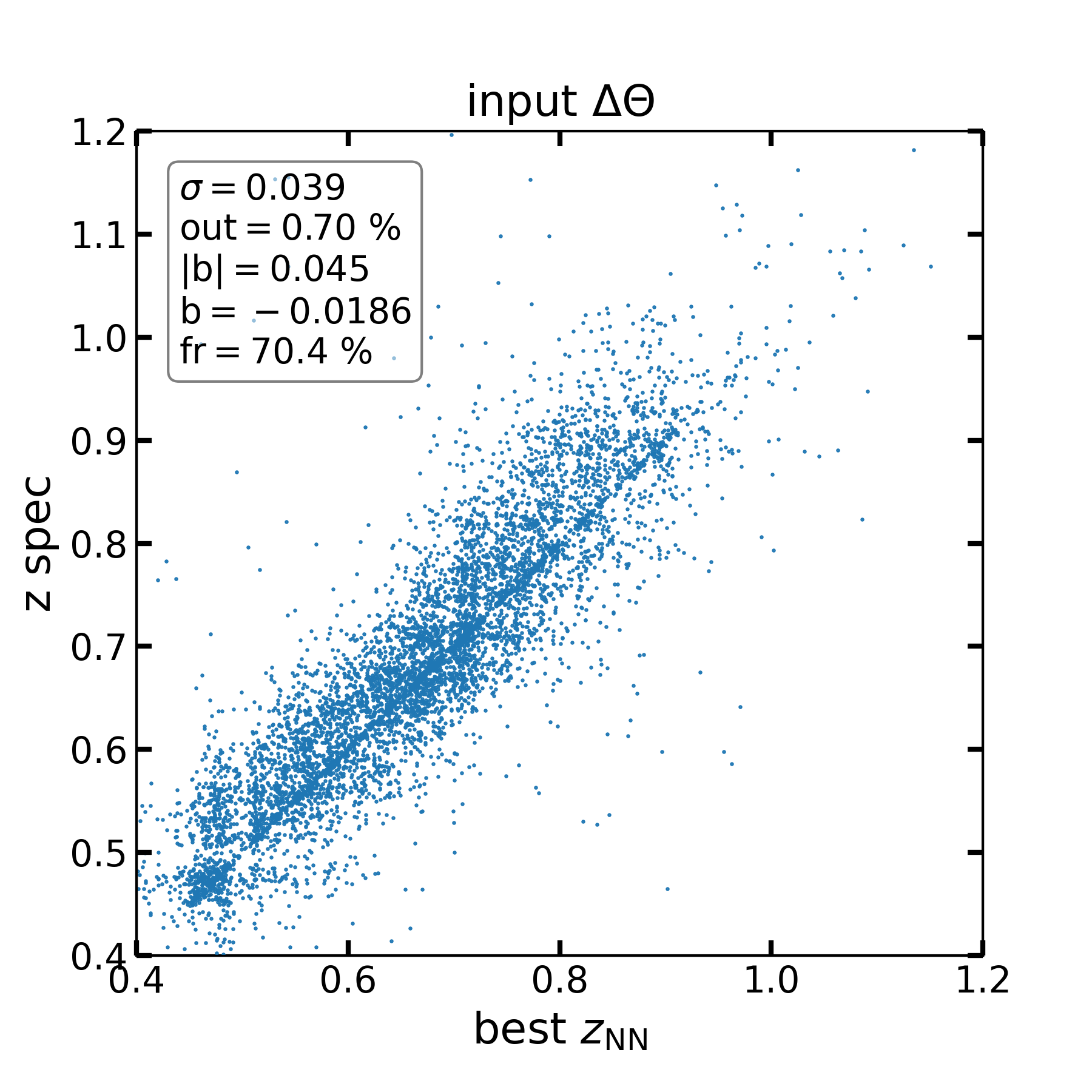}
\caption{Results of redshift estimates for the target galaxies, in the case where the angular separation Eq. (\ref{eq:angular_distance}) is an explicit input of the model. Many galaxies have slightly lower values than the real spectroscopic value, resulting in a large bias $b$. Currently, we do not have an explanation of this observed effect.
\label{fig:theta}}
\end{figure}

As a further test, we also computed the gradients of the predictions with respect to their input variables to detect the most relevant ones, as shown in Fig. \ref{fig:gradients}. It is interesting to see that the neighbour redshift is a relevant input, as expected, and some of the photometric bands are even more relevant. This confirms the intuition that the photometric information of the neighbours does indeed provide additional information about the relative distance from the target. In this plot, we also show results for the case when the angular separation is considered as one of the input variables. These results show that the angular separation $\Delta \Theta$ between the target and the neighbour does affect the predictions. This manifests itself as a bias in the redshift estimates, as visible in Fig. \ref{fig:theta}: in this case, NezNet systematically favours neighbours that are closer to us than the target, increasing the value of the bias $b$ (Eq. \ref{eq:bias}). We also tested what happens when the angular separation information is rather given in terms of the relative difference in the angular coordinates ${\rm RA}$ and ${\rm Dec}$ of the two galaxies. In this case, the bias disappears and the results are comparable to the standard case in which no angle information is provided. However, in this case, the two parameters clearly have smaller gradients than when $\Delta \Theta$ alone is considered, which suggests that they do in fact not contribute to the predicting power of the model. For these reasons, the angular separation is not considered as input variable in our final results.

\begin{figure}
\centering
\includegraphics[width=.5\textwidth]{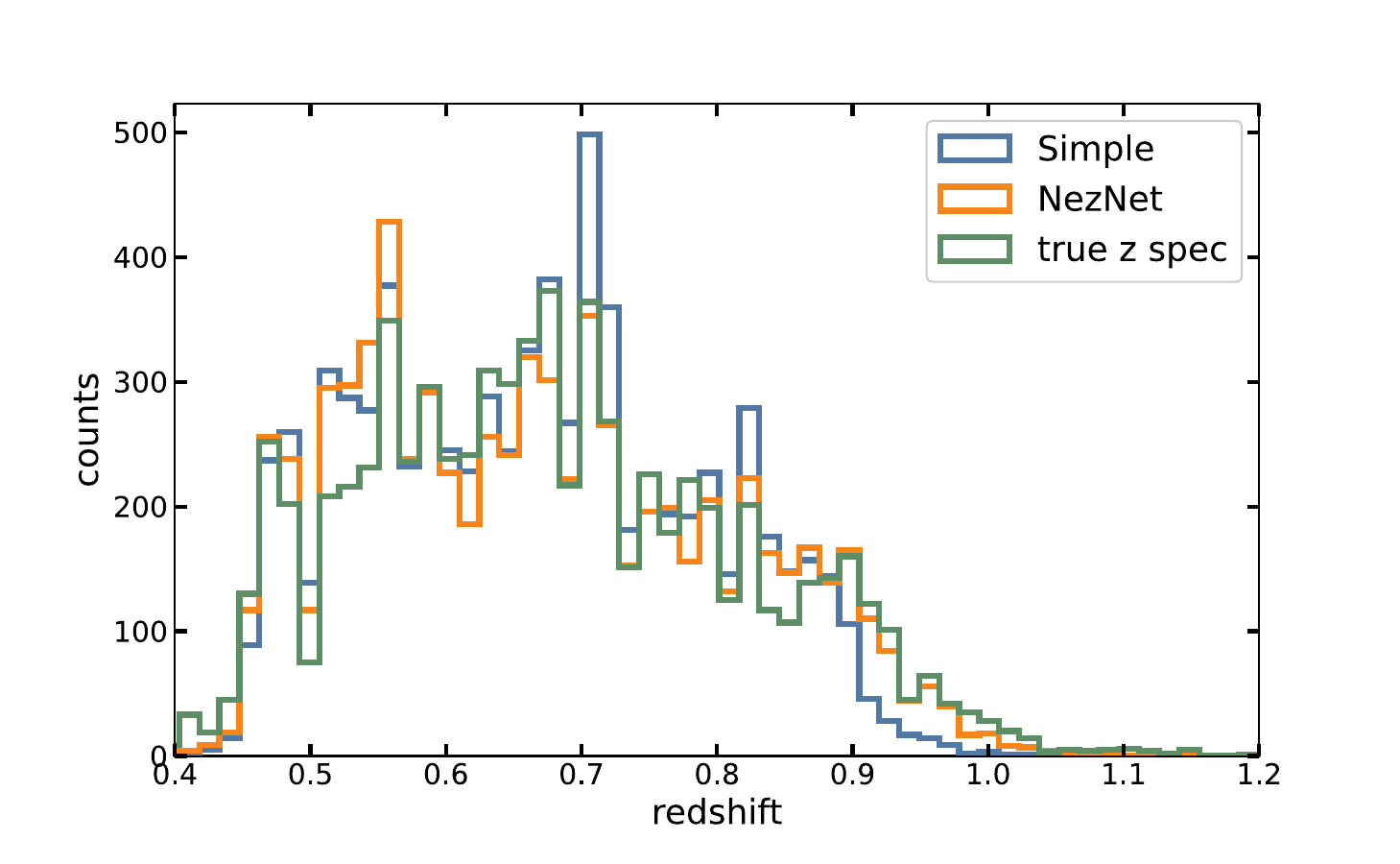}
\caption{Comparison of the redshift distribution for the predictions of NezNet, and a simpler graph model without message passing. While the latter performs reasonably well in general, it tends to cut the tail of the distribution. \label{fig:PDFs}}
\end{figure}
One of the novelties of NezNet is the message passing between node features. This is where GNNs differ from a standard ANN, where all input variables of both galaxies would be provided directly to dense layers. We also experimented with a simpler graph model, closely resembling the architecture of NezNet, but without message passing. The input features were processed independently by MLP layers for each node (we tried using either just one or several layers). The new architecture is as in Figure \ref{fig:model}, with the exception of $\textbf{h}$ function blocks, which are now substituted with new MLP blocks, without applying any message passing. The $\pmb{x}'_i$ features are summed by the aggregation function, and the summed features are mapped to the output probability through final dense layers with sigmoid activation output, just like in the model with message passing. This kind of model, which maintains the permutation invariance property of a graph, is often referred to as a deep set \citep{Zaheer17}. We find that this simple model still works remarkably well and is comparable to NezNet in general. However, it systematically cuts off the high-redshift tail of the catalogue (Fig. \ref{fig:PDFs}), even though the overall metrics remain good.

\section{Conclusions}\label{sec:conclusions}
We have presented a new ML model, dubbed NezNet, which for a pair of galaxies takes as input their measured fluxes in a number of bands together with the redshift of one of the two galaxies. NezNet is capable of probabilistically learning whether their redshift distance is below a given threshold $\Delta z$, which is set as a hyper-parameter of the model. The angular separation between the galaxies is implicit in the training set, as for every target galaxy we select its first $n_{\text{NN}}$ angular neighbours (another hyper-parameter), but it can be an explicit input variable of the model. The backbone of the model is a GNN, a class of neural networks based on message passing and the aggregation of features (Fig. \ref{fig:model}). This message passing is explicitly performed as a relative difference between features (Eq. \ref{eq:EdgeConv}).
\begin{figure*}
\centering
\includegraphics[width=1\textwidth]{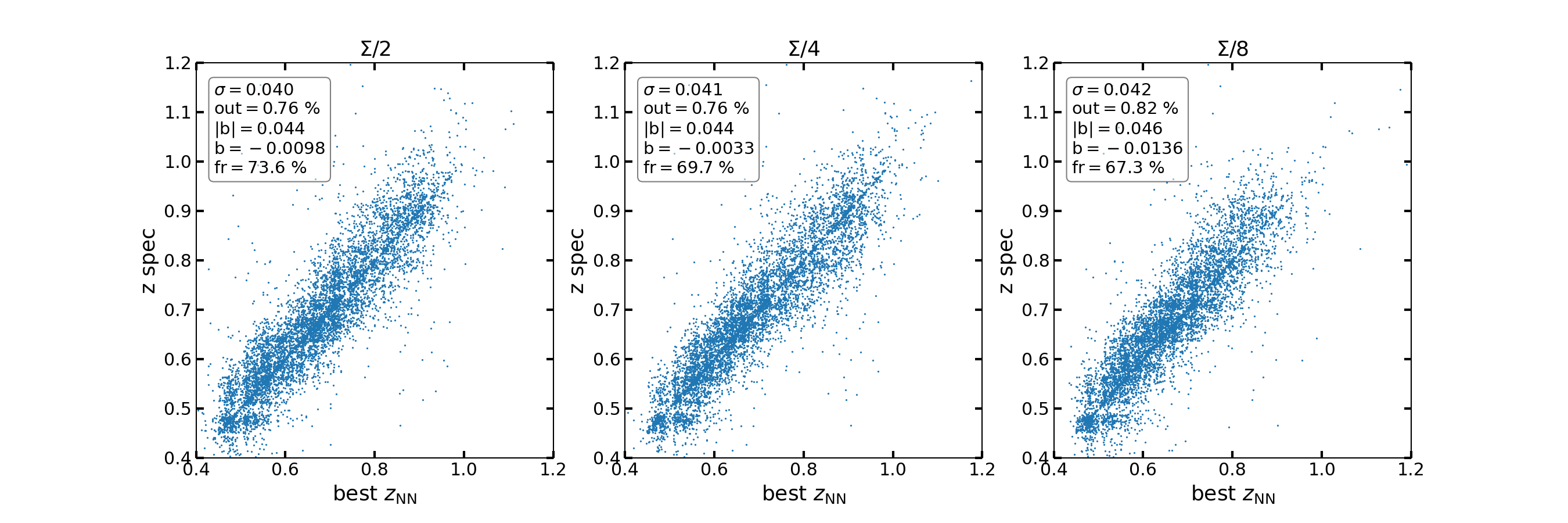}
\caption{Redshift estimates based on the best nearest neighbour, obtained by uniformly subsampling the W1 catalogue, at fixed $n_{\text{NN}}=30$ and $\Delta z= 0.08$. The titles of the panels refer to the surface density of spectroscopic objects of W1 used for training, with $\Sigma$ referring to the complete W1 sample. Except for minor fluctuations in the redshift statistics, NezNet maintains a performance similar to the case without subsampling. The only noticeable trend is the fraction of central galaxies for which a physical pair is found, which decreases for lower densities. This could be due to the decreasing number of available training data. The percentage of real physical neighbours for a central galaxy, which decreases only slightly from $\Sigma$ to $\Sigma/8$, remains around $40 \ \%$ and explains why NezNet is still effective.}
\label{fig:subsample}
\end{figure*}

NezNet outputs the score probability for a galaxy pair to be real neighbours. This information that can be used in two ways. On the one hand, if none of the $n_{\text{NN}}$ nearest neighbours is identified as a physical neighbour, the target galaxy can be considered an outlier in terms of its properties. This may suggest that it is an interloper, that is, a foreground or background object with respect to the volume sampled by the spectroscopic sample we used for the comparison. It should therefore be discarded from any sample that aims to cover the same redshift range as the spectroscopic catalogue, for instance, via photometrically estimated redshifts.  We have proved this to be true using the VIPERS catalogue. On the other hand, if a physical neighbour is identified, the target galaxy can be assigned the spectroscopic redshift of the highest scoring galaxy among the $n_{\text{NN}}$ angular neighbours, providing an independent estimate of its redshift in this way.

These results are summarised in Fig. \ref{fig:results_008} and Fig. \ref{fig:results_015}: when outliers as detected by NezNet are discarded, all the metrics of the sample improve considerably. Moreover, the NezNet redshift estimates are comparable to or superior in precision to SED-based photometric redshifts, depending on the values chosen for the hyper-parameters. Increasing $\Delta z$ increases the goodness of the classifier (Fig. \ref{fig:roc_dz}), as well as the fraction of retained data (Fig. \ref{fig:results_dz}). Changing $n_{\text{NN}}$ has a smaller impact on the classifier (Fig. \ref{fig:roc_nn}), although it significantly affects the redshift quality metrics because a large enough $n_{\text{NN}}$ improves the probability of detecting a real neighbour; a value $n_{\text{NN}} \sim 30$ is optimal in the case of VIPERS (Fig. \ref{fig:results_nn}).

It is often the case that the fraction of the parent photometric sample without a spectroscopic measurements has a higher density than the spectroscopic sample. VIPERS indeed has a spectroscopic surface density of $\Sigma \sim 6 \cdot 10^3 / \text{deg}^2$, to compare against the photometric surface density $\Sigma_{\text{ph}} \sim 45 \cdot 10^3 / \text{deg}^2$. For this reason, we tested NezNet by varying the surface density of the spectroscopic sample used during training. We achieved this by repeating the training procedure on a uniformly subsampled catalogue extracted from W1. The test was performed on W4 without any subsampling, so that we tested for the effectiveness of NezNet trained on a lower-density catalogue. Fig. \ref{fig:subsample} shows that NezNet keeps its effectiveness even when using a subsample of one-eighth of the original spectroscopic density $\Sigma$, similar to the VIPERS ratio of spectroscopic to photometric objects.

This suggests that NezNet could have an interesting potential also in the context of future experiments, such as \textit{Euclid} or the NASA \textit{Nancy Grace Roman} mission \citep{roman}. These slitless spectroscopic surveys will indeed naturally deliver overlapping photometric and spectroscopic data, which can be combined
using NezNet to improve photometric redshift estimates.

It is worth stressing that some details of the results presented here depend on the specific features of VIPERS and its parent CFHTLS photometric sample. Some of them may have been advantageous, but others could have penalised the success of the method.
For example, the slit-placement constraints in VIPERS limits the ability to target close galaxy pairs, which introduces a shadow in the layout of a VIMOS pointing \citep[see Fig. 6 of][]{Guzzo14}, and forces a lower limit in the separation of observable galaxy pairs (see Sect. \ref{sec:method}). This means that the training sample of NezNet was not ideal in our analysis because surely many of the missed angular pairs were also physical pairs. This increases our confidence in the obtained results because it shows that for samples that are characterised by small-scale incompleteness, as is typical of surveys built using fibre or multi-slit spectrographs, the method  still also delivers very useful results. In the case of the VIPERS data, an interesting exercise in this respect would be to use the data from the VLT-VIMOS Deep Survey (VVDS) \citep{VVDS} as training sample, which used the same spectrograph, but with repeated passes over the same area of  0.5 $\rm{deg}^2$ that substantially mitigate the proximity bias. We leave this exercise for a future work.

\begin{acknowledgements}
We thank Davide Bianchi for useful suggestions during the development of this work. FT and MSC are thankful to Daniele Grattarola for insightful discussions on GNNs and the use of the Spektral library. We thank the anonymous referee for his comments and suggestions. FT and LG acknowledge financial support by grant MUR PRIN 2017 ‘From Darklight to Dark Matter’, grant no. 20179P3PKJ. LG and MSC acknowledge financial support from the Italian Space Agency, ASI agreement n.
I/023/12/0. 

\end{acknowledgements}

\bibliography{aas}{}

\begin{thebibliography}{49}
\expandafter\ifx\csname natexlab\endcsname\relax\def\natexlab#1{#1}\fi

\bibitem[{{Akeson} {et~al.}(2019){Akeson}, {Armus}, {Bachelet}, {Bailey},
  {Bartusek}, {Bellini}, {Benford}, {Bennett}, {Bhattacharya}, {Bohlin},
  {Boyer}, {Bozza}, {Bryden}, {Calchi Novati}, {Carpenter}, {Casertano},
  {Choi}, {Content}, {Dayal}, {Dressler}, {Dor{\'e}}, {Fall}, {Fan}, {Fang},
  {Filippenko}, {Finkelstein}, {Foley}, {Furlanetto}, {Kalirai}, {Gaudi},
  {Gilbert}, {Girard}, {Grady}, {Greene}, {Guhathakurta}, {Heinrich},
  {Hemmati}, {Hendel}, {Henderson}, {Henning}, {Hirata}, {Ho}, {Huff},
  {Hutter}, {Jansen}, {Jha}, {Johnson}, {Jones}, {Kasdin}, {Kelly}, {Kirshner},
  {Koekemoer}, {Kruk}, {Lewis}, {Macintosh}, {Madau}, {Malhotra}, {Mandel},
  {Massara}, {Masters}, {McEnery}, {McQuinn}, {Melchior}, {Melton},
  {Mennesson}, {Peeples}, {Penny}, {Perlmutter}, {Pisani}, {Plazas}, {Poleski},
  {Postman}, {Ranc}, {Rauscher}, {Rest}, {Roberge}, {Robertson}, {Rodney},
  {Rhoads}, {Rhodes}, {Ryan}, {Sahu}, {Sand}, {Scolnic}, {Seth}, {Shvartzvald},
  {Siellez}, {Smith}, {Spergel}, {Stassun}, {Street}, {Strolger}, {Szalay},
  {Trauger}, {Troxel}, {Turnbull}, {van der Marel}, {von der Linden}, {Wang},
  {Weinberg}, {Williams}, {Windhorst}, {Wollack}, {Wu}, {Yee}, \&
  {Zimmerman}}]{roman}
{Akeson}, R., {Armus}, L., {Bachelet}, E., {et~al.} 2019,
  \href{https://ui.adsabs.harvard.edu/abs/2019arXiv190205569A}{arXiv e-prints,
  arXiv:1902.05569}

\bibitem[{{Alam} {et~al.}(2017){Alam}, {Ata}, {Bailey}, {Beutler}, {Bizyaev},
  {Blazek}, {Bolton}, {Brownstein}, {Burden}, {Chuang}, {Comparat}, {Cuesta},
  {Dawson}, {Eisenstein}, {Escoffier}, {Gil-Mar{\'\i}n}, {Grieb}, {Hand}, {Ho},
  {Kinemuchi}, {Kirkby}, {Kitaura}, {Malanushenko}, {Malanushenko}, {Maraston},
  {McBride}, {Nichol}, {Olmstead}, {Oravetz}, {Padmanabhan},
  {Palanque-Delabrouille}, {Pan}, {Pellejero-Ibanez}, {Percival}, {Petitjean},
  {Prada}, {Price-Whelan}, {Reid}, {Rodr{\'\i}guez-Torres}, {Roe}, {Ross},
  {Ross}, {Rossi}, {Rubi{\~n}o-Mart{\'\i}n}, {Saito}, {Salazar-Albornoz},
  {Samushia}, {S{\'a}nchez}, {Satpathy}, {Schlegel}, {Schneider},
  {Sc{\'o}ccola}, {Seo}, {Sheldon}, {Simmons}, {Slosar}, {Strauss}, {Swanson},
  {Thomas}, {Tinker}, {Tojeiro}, {Maga{\~n}a}, {Vazquez}, {Verde}, {Wake},
  {Wang}, {Weinberg}, {White}, {Wood-Vasey}, {Y{\`e}che}, {Zehavi}, {Zhai}, \&
  {Zhao}}]{Alam17}
{Alam}, S., {Ata}, M., {Bailey}, S., {et~al.} 2017,
  \href{http://dx.doi.org/10.1093/mnras/stx721}{\color{magenta}\mnras},
  \href{https://ui.adsabs.harvard.edu/abs/2017MNRAS.470.2617A}{470, 2617}

\bibitem[{{Alarcon} {et~al.}(2021){Alarcon}, {Gaztanaga}, {Eriksen}, {Baugh},
  {Cabayol}, {Casas}, {Carretero}, {Castander}, {De Vicente}, {Fernandez},
  {Garcia-Bellido}, {Hildebrandt}, {Hoekstra}, {Joachimi}, {Manzoni}, {Miquel},
  {Norberg}, {Padilla}, {Renard}, {Sanchez}, {Serrano}, {Sevilla-Noarbe},
  {Siudek}, \& {Tallada-Cresp{\'\i}}}]{pau}
{Alarcon}, A., {Gaztanaga}, E., {Eriksen}, M., {et~al.} 2021,
  \href{http://dx.doi.org/10.1093/mnras/staa3659}{\color{magenta}\mnras},
  \href{https://ui.adsabs.harvard.edu/abs/2021MNRAS.501.6103A}{501, 6103}

\bibitem[{{Aragon-Calvo} {et~al.}(2015){Aragon-Calvo}, {van de Weygaert},
  {Jones}, \& {Mobasher}}]{Aragon-Calvo15}
{Aragon-Calvo}, M.~A., {van de Weygaert}, R., {Jones}, B. J.~T., \& {Mobasher},
  B. 2015,
  \href{http://dx.doi.org/10.1093/mnras/stv1903}{\color{magenta}\mnras},
  \href{https://ui.adsabs.harvard.edu/abs/2015MNRAS.454..463A}{454, 463}

\bibitem[{Arnouts {et~al.}(2002)Arnouts, Moscardini, Vanzella, Colombi,
  Cristiani, Fontana, Giallongo, Matarrese, \& Saracco}]{Arnouts03}
Arnouts, S., Moscardini, L., Vanzella, E., {et~al.} 2002,
  \href{http://dx.doi.org/10.1046/j.1365-8711.2002.04988.x}{\color{magenta}Monthly
  Notices of the Royal Astronomical Society}, 329, 329

\bibitem[{{Bautista} {et~al.}(2021){Bautista}, {Paviot}, {Vargas Maga{\~n}a},
  {de la Torre}, {Fromenteau}, {Gil-Mar{\'\i}n}, {Ross}, {Burtin}, {Dawson},
  {Hou}, {Kneib}, {de Mattia}, {Percival}, {Rossi}, {Tojeiro}, {Zhao}, {Zhao},
  {Alam}, {Brownstein}, {Chapman}, {Choi}, {Chuang}, {Escoffier}, {de la
  Macorra}, {du Mas des Bourboux}, {Mohammad}, {Moon}, {M{\"u}ller},
  {Nadathur}, {Newman}, {Schneider}, {Seo}, \& {Wang}}]{Bautista21}
{Bautista}, J.~E., {Paviot}, R., {Vargas Maga{\~n}a}, M., {et~al.} 2021,
  \href{http://dx.doi.org/10.1093/mnras/staa2800}{\color{magenta}\mnras},
  \href{https://ui.adsabs.harvard.edu/abs/2021MNRAS.500..736B}{500, 736}

\bibitem[{Beck \& Sadowski(2019)}]{Beck2019}
Beck, R. \& Sadowski, P. 2019, Refined Redshift Regression in Cosmology with
  Graph Convolution Networks,
  \url{https://ml4physicalsciences.github.io/2019/files/NeurIPS_ML4PS_2019_80.pdf}

\bibitem[{{Benitez} {et~al.}(2014){Benitez}, {Dupke}, {Moles}, {Sodre},
  {Cenarro}, {Marin-Franch}, {Taylor}, {Cristobal}, {Fernandez-Soto}, {Mendes
  de Oliveira}, {Cepa-Nogue}, {Abramo}, {Alcaniz}, {Overzier},
  {Hernandez-Monteagudo}, {Alfaro}, {Kanaan}, {Carvano}, {Reis}, {Martinez
  Gonzalez}, {Ascaso}, {Ballesteros}, {Xavier}, {Varela}, {Ederoclite},
  {Vazquez Ramio}, {Broadhurst}, {Cypriano}, {Angulo}, {Diego}, {Zandivarez},
  {Diaz}, {Melchior}, {Umetsu}, {Spinelli}, {Zitrin}, {Coe}, {Yepes}, {Vielva},
  {Sahni}, {Marcos-Caballero}, {Shu Kitaura}, {Maroto}, {Masip}, {Tsujikawa},
  {Carneiro}, {Gonzalez Nuevo}, {Carvalho}, {Reboucas}, {Carvalho}, {Abdalla},
  {Bernui}, {Pigozzo}, {Ferreira}, {Chandrachani Devi}, {Bengaly}, {Campista},
  {Amorim}, {Asari}, {Bongiovanni}, {Bonoli}, {Bruzual}, {Cardiel}, {Cava},
  {Cid Fernandes}, {Coelho}, {Cortesi}, {Delgado}, {Diaz Garcia}, {Espinosa},
  {Galliano}, {Gonzalez-Serrano}, {Falcon-Barroso}, {Fritz}, {Fernandes},
  {Gorgas}, {Hoyos}, {Jimenez-Teja}, {Lopez-Aguerri}, {Lopez-San Juan},
  {Mateus}, {Molino}, {Novais}, {OMill}, {Oteo}, {Perez-Gonzalez}, {Poggianti},
  {Proctor}, {Ricciardelli}, {Sanchez-Blazquez}, {Storchi-Bergmann}, {Telles},
  {Schoennell}, {Trujillo}, {Vazdekis}, {Viironen}, {Daflon},
  {Aparicio-Villegas}, {Rocha}, {Ribeiro}, {Borges}, {Martins}, {Marcolino},
  {Martinez-Delgado}, {Perez-Torres}, {Siffert}, {Calvao}, {Sako}, {Kessler},
  {Alvarez-Candal}, {De Pra}, {Roig}, {Lazzaro}, {Gorosabel}, {Lopes de
  Oliveira}, {Lima-Neto}, {Irwin}, {Liu}, {Alvarez}, {Balmes}, {Chueca},
  {Costa-Duarte}, {da Costa}, {Dantas}, {Diaz}, {Fabregat}, {Ferrari},
  {Gavela}, {Gracia}, {Gruel}, {Gutierrez}, {Guzman}, {Hernandez-Fernandez},
  {Herranz}, {Hurtado-Gil}, {Jablonsky}, {Laporte}, {Le Tiran}, {Licandro},
  {Lima}, {Martin}, {Martinez}, {Montero}, {Penteado}, {Pereira}, {Peris},
  {Quilis}, {Sanchez-Portal}, {Soja}, {Solano}, {Torra}, \&
  {Valdivielso}}]{jpas}
{Benitez}, N., {Dupke}, R., {Moles}, M., {et~al.} 2014,
  \href{https://ui.adsabs.harvard.edu/abs/2014arXiv1403.5237B}{arXiv e-prints,
  arXiv:1403.5237}

\bibitem[{{Blake} {et~al.}(2011){Blake}, {Brough}, {Colless}, {Contreras},
  {Couch}, {Croom}, {Davis}, {Drinkwater}, {Forster}, {Gilbank}, {Gladders},
  {Glazebrook}, {Jelliffe}, {Jurek}, {Li}, {Madore}, {Martin}, {Pimbblet},
  {Poole}, {Pracy}, {Sharp}, {Wisnioski}, {Woods}, {Wyder}, \& {Yee}}]{Blake11}
{Blake}, C., {Brough}, S., {Colless}, M., {et~al.} 2011,
  \href{http://dx.doi.org/10.1111/j.1365-2966.2011.18903.x}{\color{magenta}\mnras},
  \href{https://ui.adsabs.harvard.edu/abs/2011MNRAS.415.2876B}{415, 2876}

\bibitem[{{Bolzonella} {et~al.}(2000){Bolzonella}, {Miralles}, \&
  {Pell{\'o}}}]{Bolzonella00}
{Bolzonella}, M., {Miralles}, J.~M., \& {Pell{\'o}}, R. 2000, \aap,
  \href{https://ui.adsabs.harvard.edu/abs/2000A&A...363..476B}{363, 476}

\bibitem[{{Bottini} {et~al.}(2005){Bottini}, {Garilli}, {Maccagni}, {Tresse},
  {Le Brun}, {Le F{\`e}vre}, {Picat}, {Scaramella}, {Scodeggio}, {Vettolani},
  {Zanichelli}, {Adami}, {Arnaboldi}, {Arnouts}, {Bardelli}, {Bolzonella},
  {Cappi}, {Charlot}, {Ciliegi}, {Contini}, {Foucaud}, {Franzetti}, {Guzzo},
  {Ilbert}, {Iovino}, {McCracken}, {Marano}, {Marinoni}, {Mathez}, {Mazure},
  {Meneux}, {Merighi}, {Paltani}, {Pollo}, {Pozzetti}, {Radovich}, {Zamorani},
  \& {Zucca}}]{Bottini05}
{Bottini}, D., {Garilli}, B., {Maccagni}, D., {et~al.} 2005,
  \href{http://dx.doi.org/10.1086/432150}{\color{magenta}\pasp},
  \href{https://ui.adsabs.harvard.edu/abs/2005PASP..117..996B}{117, 996}

\bibitem[{Brescia {et~al.}(2021)Brescia, Cavuoti, Razim, Amaro, Riccio, \&
  Longo}]{Brescia21}
Brescia, M., Cavuoti, S., Razim, O., {et~al.} 2021,
  \href{http://dx.doi.org/10.3389/fspas.2021.658229}{\color{magenta}Frontiers
  in Astronomy and Space Sciences}, 8, 8

\bibitem[{{Bronstein} {et~al.}(2017){Bronstein}, {Bruna}, {LeCun}, {Szlam}, \&
  {Vandergheynst}}]{Bronstein17}
{Bronstein}, M.~M., {Bruna}, J., {LeCun}, Y., {Szlam}, A., \& {Vandergheynst},
  P. 2017,
  \href{http://dx.doi.org/10.1109/MSP.2017.2693418}{\color{magenta}IEEE Signal
  Processing Magazine},
  \href{https://ui.adsabs.harvard.edu/abs/2017ISPM...34...18B}{34, 18}

\bibitem[{{Cagliari} {et~al.}(2022){Cagliari}, {Granett}, {Guzzo},
  {Bolzonella}, {Pozzetti}, {Tutusaus}, {Camera}, {Amara}, {Auricchio},
  {Bender}, {Bodendorf}, {Bonino}, {Branchini}, {Brescia}, {Capobianco},
  {Carbone}, {Carretero}, {Castander}, {Castellano}, {Cavuoti}, {Cimatti},
  {Cledassou}, {Congedo}, {Conselice}, {Conversi}, {Copin}, {Corcione},
  {Cropper}, {Degaudenzi}, {Douspis}, {Dubath}, {Dusini}, {Ealet}, {Ferriol},
  {Fourmanoit}, {Frailis}, {Franceschi}, {Franzetti}, {Garilli}, {Giocoli},
  {Grazian}, {Grupp}, {Haugan}, {Hoekstra}, {Holmes}, {Hormuth}, {Hudelot},
  {Jahnke}, {Kermiche}, {Kiessling}, {Kilbinger}, {Kitching}, {K{\"u}mmel},
  {Kunz}, {Kurki-Suonio}, {Ligori}, {Lilje}, {Lloro}, {Maiorano}, {Mansutti},
  {Marggraf}, {Markovic}, {Massey}, {Meneghetti}, {Merlin}, {Meylan},
  {Moresco}, {Moscardini}, {Niemi}, {Padilla}, {Paltani}, {Pasian}, {Pedersen},
  {Percival}, {Pettorino}, {Pires}, {Poncet}, {Popa}, {Raison}, {Rebolo},
  {Rhodes}, {Rix}, {Roncarelli}, {Rossetti}, {Saglia}, {Scaramella},
  {Schneider}, {Scodeggio}, {Secroun}, {Seidel}, {Serrano}, {Sirignano},
  {Sirri}, {Tavagnacco}, {Taylor}, {Tereno}, {Toledo-Moreo}, {Valentijn},
  {Valenziano}, {Wang}, {Welikala}, {Weller}, {Zamorani}, {Zoubian}, {Baldi},
  {Farinelli}, {Medinaceli}, {Mei}, {Polenta}, {Romelli}, {Vassallo}, \&
  {Humphrey}}]{Cagliari2022}
{Cagliari}, M.~S., {Granett}, B.~R., {Guzzo}, L., {et~al.} 2022,
  \href{http://dx.doi.org/10.1051/0004-6361/202142224}{\color{magenta}\aap},
  \href{https://ui.adsabs.harvard.edu/abs/2022A&A...660A...9C}{660, A9}

\bibitem[{{Carliles} {et~al.}(2010){Carliles}, {Budav{\'a}ri}, {Heinis},
  {Priebe}, \& {Szalay}}]{Carliles10}
{Carliles}, S., {Budav{\'a}ri}, T., {Heinis}, S., {Priebe}, C., \& {Szalay},
  A.~S. 2010,
  \href{http://dx.doi.org/10.1088/0004-637X/712/1/511}{\color{magenta}\apj},
  \href{https://ui.adsabs.harvard.edu/abs/2010ApJ...712..511C}{712, 511}

\bibitem[{{Colless} {et~al.}(2003){Colless}, {Peterson}, {Jackson}, {Peacock},
  {Cole}, {Norberg}, {Baldry}, {Baugh}, {Bland-Hawthorn}, {Bridges}, {Cannon},
  {Collins}, {Couch}, {Cross}, {Dalton}, {De Propris}, {Driver}, {Efstathiou},
  {Ellis}, {Frenk}, {Glazebrook}, {Lahav}, {Lewis}, {Lumsden}, {Maddox},
  {Madgwick}, {Sutherland}, \& {Taylor}}]{Colless03}
{Colless}, M., {Peterson}, B.~A., {Jackson}, C., {et~al.} 2003,
  \href{https://ui.adsabs.harvard.edu/abs/2003astro.ph..6581C}{arXiv e-prints,
  astro}

\bibitem[{{Collister} \& {Lahav}(2004)}]{Collister04}
{Collister}, A.~A. \& {Lahav}, O. 2004,
  \href{http://dx.doi.org/10.1086/383254}{\color{magenta}\pasp},
  \href{https://ui.adsabs.harvard.edu/abs/2004PASP..116..345C}{116, 345}

\bibitem[{{Cucciati} {et~al.}(2014){Cucciati}, {Granett}, {Branchini},
  {Marulli}, {Iovino}, {Moscardini}, {Bel}, {Cappi}, {Peacock}, {de la Torre},
  {Bolzonella}, {Guzzo}, {Polletta}, {Fritz}, {Adami}, {Bottini}, {Coupon},
  {Davidzon}, {Franzetti}, {Fumana}, {Garilli}, {Krywult}, {Ma{\l}ek},
  {Paioro}, {Pollo}, {Scodeggio}, {Tasca}, {Vergani}, {Zanichelli}, {Di Porto},
  \& {Zamorani}}]{Cucciati2014}
{Cucciati}, O., {Granett}, B.~R., {Branchini}, E., {et~al.} 2014,
  \href{http://dx.doi.org/10.1051/0004-6361/201423409}{\color{magenta}\aap},
  \href{https://ui.adsabs.harvard.edu/abs/2014A&A...565A..67C}{565, A67}

\bibitem[{{de la Torre} {et~al.}(2017){de la Torre}, {Jullo}, {Giocoli},
  {Pezzotta}, {Bel}, {Granett}, {Guzzo}, {Garilli}, {Scodeggio}, {Bolzonella},
  {Abbas}, {Adami}, {Bottini}, {Cappi}, {Cucciati}, {Davidzon}, {Franzetti},
  {Fritz}, {Iovino}, {Krywult}, {Le Brun}, {Le F{\`e}vre}, {Maccagni},
  {Ma{\l}ek}, {Marulli}, {Polletta}, {Pollo}, {Tasca}, {Tojeiro}, {Vergani},
  {Zanichelli}, {Arnouts}, {Branchini}, {Coupon}, {De Lucia}, {Ilbert},
  {Moutard}, {Moscardini}, {Peacock}, {Metcalf}, {Prada}, \& {Yepes}}]{Torre17}
{de la Torre}, S., {Jullo}, E., {Giocoli}, C., {et~al.} 2017,
  \href{http://dx.doi.org/10.1051/0004-6361/201630276}{\color{magenta}\aap},
  \href{https://ui.adsabs.harvard.edu/abs/2017A&A...608A..44D}{608, A44}

\bibitem[{{DESI Collaboration} {et~al.}(2016){DESI Collaboration}, {Aghamousa},
  {Aguilar}, {Ahlen}, {Alam}, {Allen}, {Allende Prieto}, {Annis}, {Bailey},
  {Balland}, {Ballester}, {Baltay}, {Beaufore}, {Bebek}, {Beers}, {Bell},
  {Bernal}, {Besuner}, {Beutler}, {Blake}, {Bleuler}, {Blomqvist}, {Blum},
  {Bolton}, {Briceno}, {Brooks}, {Brownstein}, {Buckley-Geer}, {Burden},
  {Burtin}, {Busca}, {Cahn}, {Cai}, {Cardiel-Sas}, {Carlberg}, {Carton},
  {Casas}, {Castander}, {Cervantes-Cota}, {Claybaugh}, {Close}, {Coker},
  {Cole}, {Comparat}, {Cooper}, {Cousinou}, {Crocce}, {Cuby}, {Cunningham},
  {Davis}, {Dawson}, {de la Macorra}, {De Vicente}, {Delubac}, {Derwent},
  {Dey}, {Dhungana}, {Ding}, {Doel}, {Duan}, {Ealet}, {Edelstein},
  {Eftekharzadeh}, {Eisenstein}, {Elliott}, {Escoffier}, {Evatt}, {Fagrelius},
  {Fan}, {Fanning}, {Farahi}, {Farihi}, {Favole}, {Feng}, {Fernandez},
  {Findlay}, {Finkbeiner}, {Fitzpatrick}, {Flaugher}, {Flender}, {Font-Ribera},
  {Forero-Romero}, {Fosalba}, {Frenk}, {Fumagalli}, {Gaensicke}, {Gallo},
  {Garcia-Bellido}, {Gaztanaga}, {Pietro Gentile Fusillo}, {Gerard},
  {Gershkovich}, {Giannantonio}, {Gillet}, {Gonzalez-de-Rivera},
  {Gonzalez-Perez}, {Gott}, {Graur}, {Gutierrez}, {Guy}, {Habib}, {Heetderks},
  {Heetderks}, {Heitmann}, {Hellwing}, {Herrera}, {Ho}, {Holland}, {Honscheid},
  {Huff}, {Hutchinson}, {Huterer}, {Hwang}, {Illa Laguna}, {Ishikawa},
  {Jacobs}, {Jeffrey}, {Jelinsky}, {Jennings}, {Jiang}, {Jimenez}, {Johnson},
  {Joyce}, {Jullo}, {Juneau}, {Kama}, {Karcher}, {Karkar}, {Kehoe}, {Kennamer},
  {Kent}, {Kilbinger}, {Kim}, {Kirkby}, {Kisner}, {Kitanidis}, {Kneib},
  {Koposov}, {Kovacs}, {Koyama}, {Kremin}, {Kron}, {Kronig}, {Kueter-Young},
  {Lacey}, {Lafever}, {Lahav}, {Lambert}, {Lampton}, {Landriau}, {Lang},
  {Lauer}, {Le Goff}, {Le Guillou}, {Le Van Suu}, {Lee}, {Lee}, {Leitner},
  {Lesser}, {Levi}, {L'Huillier}, {Li}, {Liang}, {Lin}, {Linder}, {Loebman},
  {Luki{\'c}}, {Ma}, {MacCrann}, {Magneville}, {Makarem}, {Manera}, {Manser},
  {Marshall}, {Martini}, {Massey}, {Matheson}, {McCauley}, {McDonald},
  {McGreer}, {Meisner}, {Metcalfe}, {Miller}, {Miquel}, {Moustakas}, {Myers},
  {Naik}, {Newman}, {Nichol}, {Nicola}, {Nicolati da Costa}, {Nie}, {Niz},
  {Norberg}, {Nord}, {Norman}, {Nugent}, {O'Brien}, {Oh}, {Olsen}, {Padilla},
  {Padmanabhan}, {Padmanabhan}, {Palanque-Delabrouille}, {Palmese},
  {Pappalardo}, {P{\^a}ris}, {Park}, {Patej}, {Peacock}, {Peiris}, {Peng},
  {Percival}, {Perruchot}, {Pieri}, {Pogge}, {Pollack}, {Poppett}, {Prada},
  {Prakash}, {Probst}, {Rabinowitz}, {Raichoor}, {Ree}, {Refregier}, {Regal},
  {Reid}, {Reil}, {Rezaie}, {Rockosi}, {Roe}, {Ronayette}, {Roodman}, {Ross},
  {Ross}, {Rossi}, {Rozo}, {Ruhlmann-Kleider}, {Rykoff}, {Sabiu}, {Samushia},
  {Sanchez}, {Sanchez}, {Schlegel}, {Schneider}, {Schubnell}, {Secroun},
  {Seljak}, {Seo}, {Serrano}, {Shafieloo}, {Shan}, {Sharples}, {Sholl},
  {Shourt}, {Silber}, {Silva}, {Sirk}, {Slosar}, {Smith}, {Smoot}, {Som},
  {Song}, {Sprayberry}, {Staten}, {Stefanik}, {Tarle}, {Sien Tie}, {Tinker},
  {Tojeiro}, {Valdes}, {Valenzuela}, {Valluri}, {Vargas-Magana}, {Verde},
  {Walker}, {Wang}, {Wang}, {Weaver}, {Weaverdyck}, {Wechsler}, {Weinberg},
  {White}, {Yang}, {Yeche}, {Zhang}, {Zhao}, {Zheng}, {Zhou}, {Zhou}, {Zhu},
  {Zou}, \& {Zu}}]{DESI16}
{DESI Collaboration}, {Aghamousa}, A., {Aguilar}, J., {et~al.} 2016,
  \href{https://ui.adsabs.harvard.edu/abs/2016arXiv161100036D}{arXiv e-prints,
  arXiv:1611.00036}

\bibitem[{{Gilmer} {et~al.}(2017){Gilmer}, {Schoenholz}, {Riley}, {Vinyals}, \&
  {Dahl}}]{Gilmer17}
{Gilmer}, J., {Schoenholz}, S.~S., {Riley}, P.~F., {Vinyals}, O., \& {Dahl},
  G.~E. 2017,
  \href{https://ui.adsabs.harvard.edu/abs/2017arXiv170401212G}{arXiv e-prints,
  arXiv:1704.01212}

\bibitem[{{Grattarola} \& {Alippi}(2020)}]{Grattarola20}
{Grattarola}, D. \& {Alippi}, C. 2020,
  \href{https://ui.adsabs.harvard.edu/abs/2020arXiv200612138G}{arXiv e-prints,
  arXiv:2006.12138}

\bibitem[{{Guzzo} {et~al.}(2014){Guzzo}, {Scodeggio}, {Garilli}, {Granett},
  {Fritz}, {Abbas}, {Adami}, {Arnouts}, {Bel}, {Bolzonella}, {Bottini},
  {Branchini}, {Cappi}, {Coupon}, {Cucciati}, {Davidzon}, {De Lucia}, {de la
  Torre}, {Franzetti}, {Fumana}, {Hudelot}, {Ilbert}, {Iovino}, {Krywult}, {Le
  Brun}, {Le F{\`e}vre}, {Maccagni}, {Ma{\l}ek}, {Marulli}, {McCracken},
  {Paioro}, {Peacock}, {Polletta}, {Pollo}, {Schlagenhaufer}, {Tasca},
  {Tojeiro}, {Vergani}, {Zamorani}, {Zanichelli}, {Burden}, {Di Porto},
  {Marchetti}, {Marinoni}, {Mellier}, {Moscardini}, {Nichol}, {Percival},
  {Phleps}, \& {Wolk}}]{Guzzo14}
{Guzzo}, L., {Scodeggio}, M., {Garilli}, B., {et~al.} 2014,
  \href{http://dx.doi.org/10.1051/0004-6361/201321489}{\color{magenta}\aap},
  \href{https://ui.adsabs.harvard.edu/abs/2014A&A...566A.108G}{566, A108}

\bibitem[{{Henghes} {et~al.}(2022){Henghes}, {Thiyagalingam}, {Pettitt}, {Hey},
  \& {Lahav}}]{Henghes22}
{Henghes}, B., {Thiyagalingam}, J., {Pettitt}, C., {Hey}, T., \& {Lahav}, O.
  2022, \href{http://dx.doi.org/10.1093/mnras/stac480}{\color{magenta}\mnras},
  \href{https://ui.adsabs.harvard.edu/abs/2022MNRAS.512.1696H}{512, 1696}

\bibitem[{{Ilbert} {et~al.}(2006){Ilbert}, {Arnouts}, {McCracken},
  {Bolzonella}, {Bertin}, {Le F{\`e}vre}, {Mellier}, {Zamorani}, {Pell{\`o}},
  {Iovino}, {Tresse}, {Le Brun}, {Bottini}, {Garilli}, {Maccagni}, {Picat},
  {Scaramella}, {Scodeggio}, {Vettolani}, {Zanichelli}, {Adami}, {Bardelli},
  {Cappi}, {Charlot}, {Ciliegi}, {Contini}, {Cucciati}, {Foucaud}, {Franzetti},
  {Gavignaud}, {Guzzo}, {Marano}, {Marinoni}, {Mazure}, {Meneux}, {Merighi},
  {Paltani}, {Pollo}, {Pozzetti}, {Radovich}, {Zucca}, {Bondi}, {Bongiorno},
  {Busarello}, {de La Torre}, {Gregorini}, {Lamareille}, {Mathez}, {Merluzzi},
  {Ripepi}, {Rizzo}, \& {Vergani}}]{Ilbert06}
{Ilbert}, O., {Arnouts}, S., {McCracken}, H.~J., {et~al.} 2006,
  \href{http://dx.doi.org/10.1051/0004-6361:20065138}{\color{magenta}\aap},
  \href{https://ui.adsabs.harvard.edu/abs/2006A&A...457..841I}{457, 841}

\bibitem[{Lahav(1994)}]{Lahav94}
Lahav, O. 1994,
  \href{http://dx.doi.org/https://doi.org/10.1016/0083-6656(94)90034-5}{\color{magenta}Vistas
  in Astronomy}, 38, 38

\bibitem[{{Laigle} {et~al.}(2016){Laigle}, {McCracken}, {Ilbert}, {Hsieh},
  {Davidzon}, {Capak}, {Hasinger}, {Silverman}, {Pichon}, {Coupon}, {Aussel},
  {Le Borgne}, {Caputi}, {Cassata}, {Chang}, {Civano}, {Dunlop}, {Fynbo},
  {Kartaltepe}, {Koekemoer}, {Le F{\`e}vre}, {Le Floc'h}, {Leauthaud}, {Lilly},
  {Lin}, {Marchesi}, {Milvang-Jensen}, {Salvato}, {Sanders}, {Scoville},
  {Smolcic}, {Stockmann}, {Taniguchi}, {Tasca}, {Toft}, {Vaccari}, \&
  {Zabl}}]{cosmos}
{Laigle}, C., {McCracken}, H.~J., {Ilbert}, O., {et~al.} 2016,
  \href{http://dx.doi.org/10.3847/0067-0049/224/2/24}{\color{magenta}\apjs},
  \href{https://ui.adsabs.harvard.edu/abs/2016ApJS..224...24L}{224, 24}

\bibitem[{{Laureijs} {et~al.}(2011){Laureijs}, {Amiaux}, {Arduini},
  {Augu{\`e}res}, {Brinchmann}, {Cole}, {Cropper}, {Dabin}, {Duvet}, {Ealet},
  {Garilli}, {Gondoin}, {Guzzo}, {Hoar}, {Hoekstra}, {Holmes}, {Kitching},
  {Maciaszek}, {Mellier}, {Pasian}, {Percival}, {Rhodes}, {Saavedra Criado},
  {Sauvage}, {Scaramella}, {Valenziano}, {Warren}, {Bender}, {Castander},
  {Cimatti}, {Le F{\`e}vre}, {Kurki-Suonio}, {Levi}, {Lilje}, {Meylan},
  {Nichol}, {Pedersen}, {Popa}, {Rebolo Lopez}, {Rix}, {Rottgering},
  {Zeilinger}, {Grupp}, {Hudelot}, {Massey}, {Meneghetti}, {Miller}, {Paltani},
  {Paulin-Henriksson}, {Pires}, {Saxton}, {Schrabback}, {Seidel}, {Walsh},
  {Aghanim}, {Amendola}, {Bartlett}, {Baccigalupi}, {Beaulieu}, {Benabed},
  {Cuby}, {Elbaz}, {Fosalba}, {Gavazzi}, {Helmi}, {Hook}, {Irwin}, {Kneib},
  {Kunz}, {Mannucci}, {Moscardini}, {Tao}, {Teyssier}, {Weller}, {Zamorani},
  {Zapatero Osorio}, {Boulade}, {Foumond}, {Di Giorgio}, {Guttridge}, {James},
  {Kemp}, {Martignac}, {Spencer}, {Walton}, {Bl{\"u}mchen}, {Bonoli},
  {Bortoletto}, {Cerna}, {Corcione}, {Fabron}, {Jahnke}, {Ligori}, {Madrid},
  {Martin}, {Morgante}, {Pamplona}, {Prieto}, {Riva}, {Toledo}, {Trifoglio},
  {Zerbi}, {Abdalla}, {Douspis}, {Grenet}, {Borgani}, {Bouwens}, {Courbin},
  {Delouis}, {Dubath}, {Fontana}, {Frailis}, {Grazian}, {Koppenh{\"o}fer},
  {Mansutti}, {Melchior}, {Mignoli}, {Mohr}, {Neissner}, {Noddle}, {Poncet},
  {Scodeggio}, {Serrano}, {Shane}, {Starck}, {Surace}, {Taylor},
  {Verdoes-Kleijn}, {Vuerli}, {Williams}, {Zacchei}, {Altieri}, {Escudero
  Sanz}, {Kohley}, {Oosterbroek}, {Astier}, {Bacon}, {Bardelli}, {Baugh},
  {Bellagamba}, {Benoist}, {Bianchi}, {Biviano}, {Branchini}, {Carbone},
  {Cardone}, {Clements}, {Colombi}, {Conselice}, {Cresci}, {Deacon}, {Dunlop},
  {Fedeli}, {Fontanot}, {Franzetti}, {Giocoli}, {Garcia-Bellido}, {Gow},
  {Heavens}, {Hewett}, {Heymans}, {Holland}, {Huang}, {Ilbert}, {Joachimi},
  {Jennins}, {Kerins}, {Kiessling}, {Kirk}, {Kotak}, {Krause}, {Lahav}, {van
  Leeuwen}, {Lesgourgues}, {Lombardi}, {Magliocchetti}, {Maguire}, {Majerotto},
  {Maoli}, {Marulli}, {Maurogordato}, {McCracken}, {McLure}, {Melchiorri},
  {Merson}, {Moresco}, {Nonino}, {Norberg}, {Peacock}, {Pello}, {Penny},
  {Pettorino}, {Di Porto}, {Pozzetti}, {Quercellini}, {Radovich}, {Rassat},
  {Roche}, {Ronayette}, {Rossetti}, {Sartoris}, {Schneider}, {Semboloni},
  {Serjeant}, {Simpson}, {Skordis}, {Smadja}, {Smartt}, {Spano}, {Spiro},
  {Sullivan}, {Tilquin}, {Trotta}, {Verde}, {Wang}, {Williger}, {Zhao},
  {Zoubian}, \& {Zucca}}]{Laureijs}
{Laureijs}, R., {Amiaux}, J., {Arduini}, S., {et~al.} 2011,
  \href{https://ui.adsabs.harvard.edu/abs/2011arXiv1110.3193L}{arXiv e-prints,
  arXiv:1110.3193}

\bibitem[{{Le F{\`e}vre} {et~al.}(2005){Le F{\`e}vre}, {Vettolani}, {Garilli},
  {Tresse}, {Bottini}, {Le Brun}, {Maccagni}, {Picat}, {Scaramella},
  {Scodeggio}, {Zanichelli}, {Adami}, {Arnaboldi}, {Arnouts}, {Bardelli},
  {Bolzonella}, {Cappi}, {Charlot}, {Ciliegi}, {Contini}, {Foucaud},
  {Franzetti}, {Gavignaud}, {Guzzo}, {Ilbert}, {Iovino}, {McCracken}, {Marano},
  {Marinoni}, {Mathez}, {Mazure}, {Meneux}, {Merighi}, {Paltani}, {Pell{\`o}},
  {Pollo}, {Pozzetti}, {Radovich}, {Zamorani}, {Zucca}, {Bondi}, {Bongiorno},
  {Busarello}, {Lamareille}, {Mellier}, {Merluzzi}, {Ripepi}, \&
  {Rizzo}}]{VVDS}
{Le F{\`e}vre}, O., {Vettolani}, G., {Garilli}, B., {et~al.} 2005,
  \href{http://dx.doi.org/10.1051/0004-6361:20041960}{\color{magenta}\aap},
  \href{https://ui.adsabs.harvard.edu/abs/2005A&A...439..845L}{439, 845}

\bibitem[{LeCun {et~al.}(2015)LeCun, Bengio, \& Hinton}]{LeCun15}
LeCun, Y., Bengio, Y., \& Hinton, G. 2015, nature, 521, 521

\bibitem[{{Maraston}(2005)}]{Maraston05}
{Maraston}, C. 2005,
  \href{http://dx.doi.org/10.1111/j.1365-2966.2005.09270.x}{\color{magenta}\mnras},
  \href{https://ui.adsabs.harvard.edu/abs/2005MNRAS.362..799M}{362, 799}

\bibitem[{Masters {et~al.}(2015)Masters, Capak, Stern, Ilbert, Salvato,
  Schmidt, Longo, Rhodes, Paltani, Mobasher, Hoekstra, Hildebrandt, Coupon,
  Steinhardt, Speagle, Faisst, Kalinich, Brodwin, Brescia, \&
  Cavuoti}]{Masters15}
Masters, D., Capak, P., Stern, D., {et~al.} 2015,
  \href{http://dx.doi.org/10.1088/0004-637x/813/1/53}{\color{magenta}The
  Astrophysical Journal}, 813, 813

\bibitem[{{Moutard} {et~al.}(2016){Moutard}, {Arnouts}, {Ilbert}, {Coupon},
  {Hudelot}, {Vibert}, {Comte}, {Conseil}, {Davidzon}, {Guzzo}, {Llebaria},
  {Martin}, {McCracken}, {Milliard}, {Morrison}, {Schiminovich}, {Treyer}, \&
  {Van Werbaeke}}]{Moutard16}
{Moutard}, T., {Arnouts}, S., {Ilbert}, O., {et~al.} 2016,
  \href{http://dx.doi.org/10.1051/0004-6361/201527945}{\color{magenta}\aap},
  \href{https://ui.adsabs.harvard.edu/abs/2016A&A...590A.102M}{590, A102}

\bibitem[{Murtagh(1991)}]{Murtagh91}
Murtagh, F. 1991,
  \href{http://dx.doi.org/https://doi.org/10.1016/0925-2312(91)90023-5}{\color{magenta}Neurocomputing},
  2, 2

\bibitem[{{Newman}(2008)}]{Newman08}
{Newman}, J.~A. 2008,
  \href{http://dx.doi.org/10.1086/589982}{\color{magenta}\apj},
  \href{https://ui.adsabs.harvard.edu/abs/2008ApJ...684...88N}{684, 88}

\bibitem[{Newman \& Gruen(2022)}]{Newman22}
Newman, J.~A. \& Gruen, D. 2022,
  \href{http://dx.doi.org/10.1146/annurev-astro-032122-014611}{\color{magenta}Annual
  Review of Astronomy and Astrophysics}, 60, 60

\bibitem[{{O'Shea} \& {Nash}(2015)}]{OShea15}
{O'Shea}, K. \& {Nash}, R. 2015,
  \href{https://ui.adsabs.harvard.edu/abs/2015arXiv151108458O}{arXiv e-prints,
  arXiv:1511.08458}

\bibitem[{{Pasquet} {et~al.}(2019){Pasquet}, {Bertin}, {Treyer}, {Arnouts}, \&
  {Fouchez}}]{Pasquet19}
{Pasquet}, J., {Bertin}, E., {Treyer}, M., {Arnouts}, S., \& {Fouchez}, D.
  2019,
  \href{http://dx.doi.org/10.1051/0004-6361/201833617}{\color{magenta}\aap},
  \href{https://ui.adsabs.harvard.edu/abs/2019A&A...621A..26P}{621, A26}

\bibitem[{{Perlmutter} {et~al.}(1998){Perlmutter}, {Aldering}, {della Valle},
  {Deustua}, {Ellis}, {Fabbro}, {Fruchter}, {Goldhaber}, {Groom}, {Hook},
  {Kim}, {Kim}, {Knop}, {Lidman}, {McMahon}, {Nugent}, {Pain}, {Panagia},
  {Pennypacker}, {Ruiz-Lapuente}, {Schaefer}, \& {Walton}}]{Perlmutter98}
{Perlmutter}, S., {Aldering}, G., {della Valle}, M., {et~al.} 1998,
  \href{http://dx.doi.org/10.1038/34124}{\color{magenta}\nat},
  \href{https://ui.adsabs.harvard.edu/abs/1998Natur.391...51P}{391, 51}

\bibitem[{{Pezzotta} {et~al.}(2017){Pezzotta}, {de la Torre}, {Bel}, {Granett},
  {Guzzo}, {Peacock}, {Garilli}, {Scodeggio}, {Bolzonella}, {Abbas}, {Adami},
  {Bottini}, {Cappi}, {Cucciati}, {Davidzon}, {Franzetti}, {Fritz}, {Iovino},
  {Krywult}, {Le Brun}, {Le F{\`e}vre}, {Maccagni}, {Ma{\l}ek}, {Marulli},
  {Polletta}, {Pollo}, {Tasca}, {Tojeiro}, {Vergani}, {Zanichelli}, {Arnouts},
  {Branchini}, {Coupon}, {De Lucia}, {Koda}, {Ilbert}, {Mohammad}, {Moutard},
  \& {Moscardini}}]{Pezzotta17}
{Pezzotta}, A., {de la Torre}, S., {Bel}, J., {et~al.} 2017,
  \href{http://dx.doi.org/10.1051/0004-6361/201630295}{\color{magenta}\aap},
  \href{https://ui.adsabs.harvard.edu/abs/2017A&A...604A..33P}{604, A33}

\bibitem[{{Riess} {et~al.}(1998){Riess}, {Filippenko}, {Challis},
  {Clocchiatti}, {Diercks}, {Garnavich}, {Gilliland}, {Hogan}, {Jha},
  {Kirshner}, {Leibundgut}, {Phillips}, {Reiss}, {Schmidt}, {Schommer},
  {Smith}, {Spyromilio}, {Stubbs}, {Suntzeff}, \& {Tonry}}]{Riess98}
{Riess}, A.~G., {Filippenko}, A.~V., {Challis}, P., {et~al.} 1998,
  \href{http://dx.doi.org/10.1086/300499}{\color{magenta}\aj},
  \href{https://ui.adsabs.harvard.edu/abs/1998AJ....116.1009R}{116, 1009}

\bibitem[{{Sadeh} {et~al.}(2016){Sadeh}, {Abdalla}, \& {Lahav}}]{Sadeh16}
{Sadeh}, I., {Abdalla}, F.~B., \& {Lahav}, O. 2016,
  \href{http://dx.doi.org/10.1088/1538-3873/128/968/104502}{\color{magenta}\pasp},
  \href{https://ui.adsabs.harvard.edu/abs/2016PASP..128j4502S}{128, 104502}

\bibitem[{{Salvato} {et~al.}(2019){Salvato}, {Ilbert}, \& {Hoyle}}]{Salvato19}
{Salvato}, M., {Ilbert}, O., \& {Hoyle}, B. 2019,
  \href{http://dx.doi.org/10.1038/s41550-018-0478-0}{\color{magenta}Nature
  Astronomy}, \href{https://ui.adsabs.harvard.edu/abs/2019NatAs...3..212S}{3,
  212}

\bibitem[{{Scodeggio} {et~al.}(2018){Scodeggio}, {Guzzo}, {Garilli}, {Granett},
  {Bolzonella}, {de la Torre}, {Abbas}, {Adami}, {Arnouts}, {Bottini}, {Cappi},
  {Coupon}, {Cucciati}, {Davidzon}, {Franzetti}, {Fritz}, {Iovino}, {Krywult},
  {Le Brun}, {Le F{\`e}vre}, {Maccagni}, {Ma{\l}ek}, {Marchetti}, {Marulli},
  {Polletta}, {Pollo}, {Tasca}, {Tojeiro}, {Vergani}, {Zanichelli}, {Bel},
  {Branchini}, {De Lucia}, {Ilbert}, {McCracken}, {Moutard}, {Peacock},
  {Zamorani}, {Burden}, {Fumana}, {Jullo}, {Marinoni}, {Mellier}, {Moscardini},
  \& {Percival}}]{Scodeggio17}
{Scodeggio}, M., {Guzzo}, L., {Garilli}, B., {et~al.} 2018,
  \href{http://dx.doi.org/10.1051/0004-6361/201630114}{\color{magenta}\aap},
  \href{https://ui.adsabs.harvard.edu/abs/2018A&A...609A..84S}{609, A84}

\bibitem[{Tegmark {et~al.}(2006)Tegmark, Eisenstein, Strauss, Weinberg,
  Blanton, Frieman, Fukugita, Gunn, Hamilton, Knapp, Nichol, Ostriker,
  Padmanabhan, Percival, Schlegel, Schneider, Scoccimarro, Seljak, Seo,
  Swanson, Szalay, Vogeley, Yoo, Zehavi, Abazajian, Anderson, Annis, Bahcall,
  Bassett, Berlind, Brinkmann, Budavari, Castander, Connolly, Csabai, Doi,
  Finkbeiner, Gillespie, Glazebrook, Hennessy, Hogg,
  Ivezi\ifmmode~\acute{c}\else \'{c}\fi{}, Jain, Johnston, Kent, Lamb, Lee,
  Lin, Loveday, Lupton, Munn, Pan, Park, Peoples, Pier, Pope, Richmond,
  Rockosi, Scranton, Sheth, Stebbins, Stoughton, Szapudi, Tucker, Berk, Yanny,
  \& York}]{Tegmark06}
Tegmark, M., Eisenstein, D.~J., Strauss, M.~A., {et~al.} 2006,
  \href{http://dx.doi.org/10.1103/PhysRevD.74.123507}{\color{magenta}Phys. Rev.
  D}, 74, 74

\bibitem[{{Tonry} \& {Davis}(1979)}]{Tonry79}
{Tonry}, J. \& {Davis}, M. 1979,
  \href{http://dx.doi.org/10.1086/112569}{\color{magenta}\aj},
  \href{https://ui.adsabs.harvard.edu/abs/1979AJ.....84.1511T}{84, 1511}

\bibitem[{{Wang} {et~al.}(2018){Wang}, {Sun}, {Liu}, {Sarma}, {Bronstein}, \&
  {Solomon}}]{Wang18}
{Wang}, Y., {Sun}, Y., {Liu}, Z., {et~al.} 2018,
  \href{https://ui.adsabs.harvard.edu/abs/2018arXiv180107829W}{arXiv e-prints,
  arXiv:1801.07829}

\bibitem[{{Zaheer} {et~al.}(2017){Zaheer}, {Kottur}, {Ravanbakhsh}, {Poczos},
  {Salakhutdinov}, \& {Smola}}]{Zaheer17}
{Zaheer}, M., {Kottur}, S., {Ravanbakhsh}, S., {et~al.} 2017,
  \href{https://ui.adsabs.harvard.edu/abs/2017arXiv170306114Z}{arXiv e-prints,
  arXiv:1703.06114}

\bibitem[{{Zhou} {et~al.}(2018){Zhou}, {Cui}, {Hu}, {Zhang}, {Yang}, {Liu},
  {Wang}, {Li}, \& {Sun}}]{Zhou18}
{Zhou}, J., {Cui}, G., {Hu}, S., {et~al.} 2018,
  \href{https://ui.adsabs.harvard.edu/abs/2018arXiv181208434Z}{arXiv e-prints,
  arXiv:1812.08434}

\end{thebibliography}
\bibliographystyle{aa_url}

\end{document}